\documentstyle[eqsecnum,aps]{revtex}

\input{epsf.tex}
\begin{document}
\draft
\twocolumn[\hsize\textwidth\columnwidth\hsize\csname
@twocolumnfalse\endcsname
\preprint{HEP/123-qed}
\renewcommand{\thefootnote}{\alph{footnote}} 
\title{Chaos, ergodicity, and the thermodynamics of lower-dimensional 
time-independent Hamiltonian systems}
\author{Henry E. Kandrup\footnote{Electronic address: kandrup@astro.ufl.edu}}
\address{ Department of Astronomy, Department of Physics, and
Institute for Fundamental Theory
\\
University of Florida, Gainesville, Florida 32611}
\author{Ioannis V. Sideris\footnote{Electronic address: sideris@astro.ufl.edu}}
\address{Department of Astronomy, University of Florida, Gainesville, 
Florida 32611\\}
\author{C. L. Bohn\footnote{Electronic address: clbohn@fnal.gov}}
\address{Fermilab, Batavia, Illinois 60510\\}
\date{\today}
\maketitle
\begin{abstract}
This paper uses the assumptions of ergodicity and a microcanonical distribution
to compute estimates of the largest Lyapunov exponents in lower-dimensional 
Hamiltonian systems. That the resulting estimates are in reasonable agreement
with the actual values computed numerically corroborates the intuition that
chaos in such systems can be understood as arising generically from a 
parametric instability and that this instability can be modeled by a 
stochastic-oscillator equation ({\it cf.} Casetti, Clementi, and Pettini,
Phys. Rev. {\bf E 54}, 5969 (1996)), linearised perturbations of a chaotic 
orbit 
satisfying a harmonic-oscillator equation with a randomly varying frequency.
\end{abstract}
]
\narrowtext
\section{INTRODUCTION AND MOTIVATION}
 \label{sec:level1}
By definition, Lyapunov exponents probe the average linear instability of 
chaotic orbits in an asymptotic $t\to\infty$ limit\cite{LL}. Their 
computation thus involves solving a matrix harmonic-oscillator equation with 
characteristic frequencies that vary in time. In the context 
of a geometric description -- which is convenient but by no means necessary
--, this equation can be reinterpreted as a Jacobi 
equation ({\it i.e.,} equation of geodesic deviation) for motion in an 
appropriately defined curved space, {\it e.g.,} by introducing the Eisenhart 
metric\cite{Ei}.

It has been long known\cite{H,A} that geodesic flows in a space with
everywhere negative curvature are unstable in the sense that nearby orbits
diverge exponentially; and, for this reason, there was an implicit assumption 
in much earlier work that chaos could often be understood as a manifestation
of negative curvature. However, as emphasized by Pettini\cite{Pet}, 
in many systems chaos cannot be attributed to negative curvature. In many 
cases, the average curvature is positive; and indeed, there 
are many known examples of nonintegrable Hamiltonian systems ({\it e.g.,} the 
finite order truncations of the Toda\cite{To} potential) which admit large 
measures of chaos even though the curvature is everywhere positive. The
curvature associated with the Eisenhart metric can be negative only if the
second derivative of the potential becomes negative. Instead, it would seem 
natural to understand chaos as reflecting a parametric instability. 
\par The Jacobi equation for a regular periodic orbit reduces to a 
multi-dimensional Hill equation, {\it i.e.,} a harmonic-oscillator equation
with frequencies that exhibit
a periodic time dependence. For certain amplitudes and periodicities, the 
solutions to such an equation remain bounded (or grow at most as a power law 
in time), this corresponding to stable periodic orbits. However, for other 
amplitudes and periodicities, the solutions grow exponentially, this 
corresponding instead to an unstable periodic orbit\cite{Kan}.

Since chaotic orbits are aperiodic and (in some sense) `random,' one might 
instead suppose that one can model the Jacobi equation describing a linearised 
perturbation of a chaotic orbit as a {\it stochastic} harmonic-oscillator 
equation, in which the time-dependent frequencies vary in a random fashion. 
Given this assumption, the key issue becomes one of identifying the stochastic 
process, {\it i.e.} the form of the colored noise, which can capture 
correctly solutions to the Jacobi equation. 

If, for fixed potential and energy, almost all of the 
constant energy hypersurface is chaotic, as is true generically for $D>2$ 
(provided that the energy $E$ is the only time-independent constant of the 
motion), it would seem reasonable to infer that the parameters for the 
oscillator equation should be estimatable assuming ergodicity. What this means
is that one can assume an invariant measure corresponding to a uniform
population of the constant energy hypersurface, {\it i.e.}, a microcanonical 
distribution. If, furthermore, one is concerned with comparatively 
high-dimensional systems, the computationally awkward description in terms of 
a microcanonical distribution can be replaced by a more user-friendly 
description based on a canonical distribution: In the spirit of ordinary 
thermodynamics, one can argue that the canonical and microcanonical ensembles 
should yield nearly identical results in the large $D$ limit. 

Given this logic, Casetti, Clementi, and Pettini\cite{CCP}
developed a `thermodynamic' theory of chaos which  they used to obtain very
good estimates of the values of the largest Lyapunov exponents for two well 
studied physical systems. To do this, they:
i) extracted from the full $D$-dimensional Jacobi equation an `isotropized' 
one-dimensional oscillator equation which they argued should capture the 
chaotic behavior of typical orbits; (ii) derived the statistics of their
assumed stochastic process in the context of a canonical ensemble description; 
and then (iii) showed that, for two seemingly generic models, solutions to 
the resulting equation yield reasonable estimates of the largest Lyapunov 
exponent, at least for $D>100$ or so.

An obvious question is whether this logic can also be exploited to provide 
reasonable estimates of the largest Lyapunov exponent for lower dimensional 
systems, say $D=2$ or $D=3$. As discussed in the concluding Section, there
are a variety of settings where it would be convenient if one could estimate 
these values without resorting to detailed numerical computations. Arguably,
however, this is {\it not} the most important point. Rather, the foremost
objective is to implement a simple physical picture of the origins of chaos in
lower-dimensional Hamiltonian systems. To the extent that the Casetti {\it et
al} proposal, or some variant thereof, can provide reasonable estimates of
Lyapunov exponents in these systems, one would seem justified in
visualising chaos as arising from a parametric instability manifested by a
stochastic-oscillator equation. In other words, one will have a clear new
paradigm in terms of which to interpret the origins of chaos in 
lower-dimensional Hamiltonian systems.

\section{AN ILLUSTRATIVE EXAMPLE}
The validity of the formula for the largest Lyapunov exponent derived by
Casetti {\it et al.} was tested for one simple toy model.  The model is 
motivated by recent observations of elliptical galaxies, which suggest that 
these objects can exhibit significant deviations from axisymmetry and that 
they often have a high-density cusp at their centroids, perhaps associated 
with the presence of a supermassive black hole.  The stars in a real galaxy 
populate a $6N$-dimensional phase space, with $N$ denoting the number of stars 
in a system.  Considering that fine structure due to localised irregularities 
and 
granularity will take a long time to manifest itself, it is of interest to 
model the system in terms of its coarse-grained 6-dimensional phase space in 
expectation that the time scale associated with the coarse-grained potential 
will represent the shortest time scale for macroscopic evolution.  Thus, a 
model potential for such systems comprises the sum of an anisotropic harmonic 
potential and a spherical Plummer potential:
\begin{equation}
V(x,y,z)={1\over 2}(a^{2}x^{2}+b^{2}y^{2}+c^{2}z^{2})-
{M_{BH}\over \sqrt{r^{2}+{\epsilon}^{2}}},
\label{eq:toy}
\end{equation}
with $r^{2}=x^{2}+y^{2}+z^{2}$, $a^2=1-\Delta$, $b^2=1$, and $c^2=1+\Delta$. 
$\Delta$ parameterizes the ellipsoidal geometry, and $\epsilon$ functions as 
a ``softening parameter'' which is set at $\epsilon=10^{-2}$ for numerical 
simulations.

The theory of Casetti {\it et al.}, described in Section IV A below is 
analytic, and within this formalism $\epsilon$ acts as a ``free parameter'' 
that reflects uncertainty about the detailed dynamical properties of the phase 
space.  One knows {\it a priori} that far from $M_{BH}$ the potential it is 
approximately 
quadratic in the coordinates, and close to $M_{BH}$ it is approximately 
spherically symmetric; the orbits are accordingly quasiregular in these 
regions wherein there will be almost no chaotic mixing.  The theory correctly 
predicts zero chaotic mixing in a harmonic-oscillator potential, thereby 
incorporating the former circumstance, but it also incorrectly predicts 
nonzero chaotic mixing in the spherically symmetric Plummer potential that 
dominates near $M_{BH}$.  Thus, a nonzero $\epsilon$ ``regularizes'' orbits 
near the black hole.  In view of these considerations, the value of $\epsilon$ 
used in the theory was chosen by requiring the magnitude of the harmonic 
potential to be comparable to a tenth of that of the Plummer potential at 
distances ``$r=\epsilon$'' from the centroid.  The specific choice is 
$\epsilon=0.5 M_{BH}^{1/3}$.

\begin{figure}[t]
\centering
\centerline{
        \epsfxsize=8cm
        \epsffile{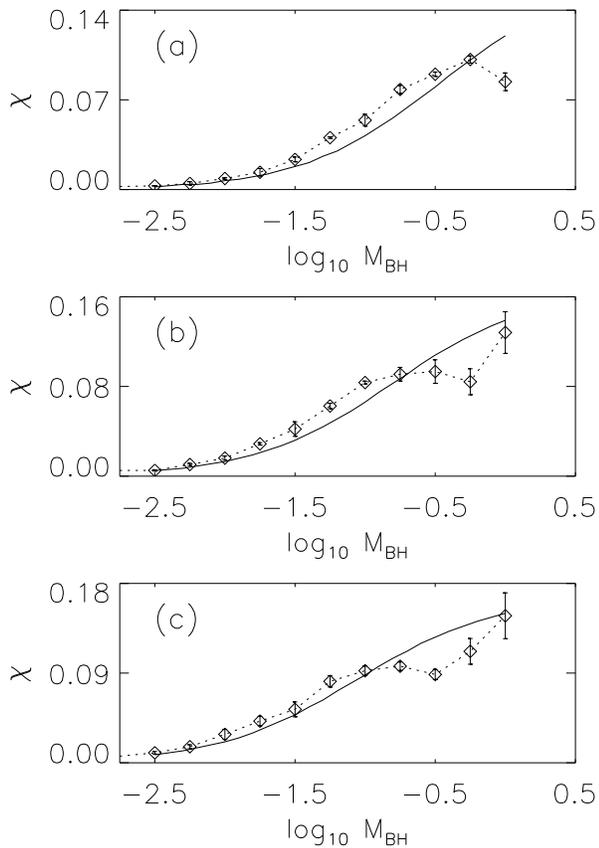}
           }
        \begin{minipage}{10cm}
        \end{minipage}
        \vskip -0.0in\hskip -0.0in
\caption{\label{fig-1} (a)
Numerical (diamonds) and analytic (solid line) estimates of the largest 
Lyapunov exponent for chaotic orbits evolved in the 3-dimensional galactic 
potential (2.1) with $a^2=0.75$, $b^2=1.0$, and $c^2=1.25$ as a function of 
black 
hole mass $M_{BH}$ for total particle energy $E=1.0$. (b) The same for
$E=0.6.$ (c) $E=0.4$.}
\end{figure}

FIGURE~\ref{fig-1} compares the ``true'' Lyapunov exponents, computed via
numerical simulations with estimates of the largest Lyapunov exponent derived 
using the Cassetti {\it et al.} formalism, Eqs. (4.12), (4.15), and (4.16) below. The numerical studies are described in detail in Ref.~\cite{kandrup01}, and the numerically generated curves derive from FIG.~5 in that paper.  The figure shows how the Lyapunov exponents pertaining to chaotic orbits scale against black-hole mass and total particle energy $E$.  Interestingly, the analytic results agree closely with the numerical results, particularly for intermediate-to-small values of $M_{BH}$. The agreement is still reasonable for large values of $M_{BH}$ (values that are in fact unphysically large), though the degree of agreement is less good. This is as expected in that a black-hole mass that is comparable to the ellipsoidal mass will establish sizeable regions of regularity over the constant-energy hypersurface, and the fraction of chaotic orbits will be correspondingly lower~\cite{kandrup01}. 

\begin{figure}[t]
\centering
\centerline{
        \epsfxsize=8cm
        \epsffile{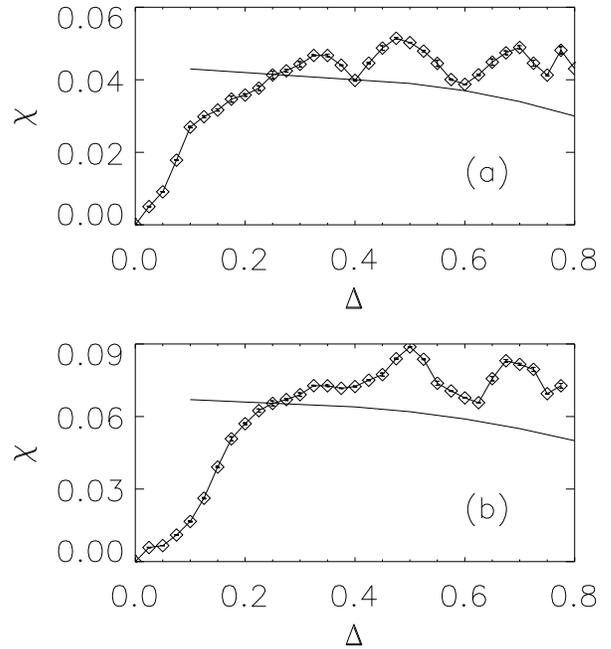}
           }
        \begin{minipage}{10cm}
        \end{minipage}
        \vskip -0.0in\hskip -0.0in
\caption{\label{fig-2} (a)
Numerical (diamonds) and analytic (solid line) estimates of the largest 
Lyapunov exponent for chaotic orbits evolved in the 3-dimensional galactic 
potential (2.1) ith $M_{BH}=0.1$ and $a^2=1-\Delta,~b^2=1,~c^2=1+\Delta$ as a 
function of $\Delta$ for total particle energy $E=1.0$. (b) The same for
$E=0.6$.}
\end{figure}

FIGURE~\ref{fig-2} compares, for fixed $M_{BH}=0.1$, the numerical and 
analytic Lyapunov exponents versus ellipticity as parameterized by $\Delta$.  
Again, the analytic technique is seen to yield reasonable estimates provided 
$\Delta$ is not too small.  As $\Delta$ decreases to zero, the potential 
approaches spherical symmetry and is thereby integrable, supporting only 
regular orbits.  Inasmuch as the fundamental assumption underlying the Casetti 
{\it et al.} formalism is that a substantial fraction of the orbits is 
globally chaotic, the formalism clearly breaks down for spherical 
symmetry. As discussed in Ref.~\cite{kandrup01}, the fact that the numerical 
curves exhibit a great deal of structure not manifested by the analytic 
predictions reflects the fact that the phase space associated with the
potential (2.1) is dominated by resonances with frequencies $a$, $b$, and $c$ 
associated with the harmonic contribution which are completely independent
of initial conditions.

\begin{figure}[t]
\centering
\centerline{
        \epsfxsize=8cm
        \epsffile{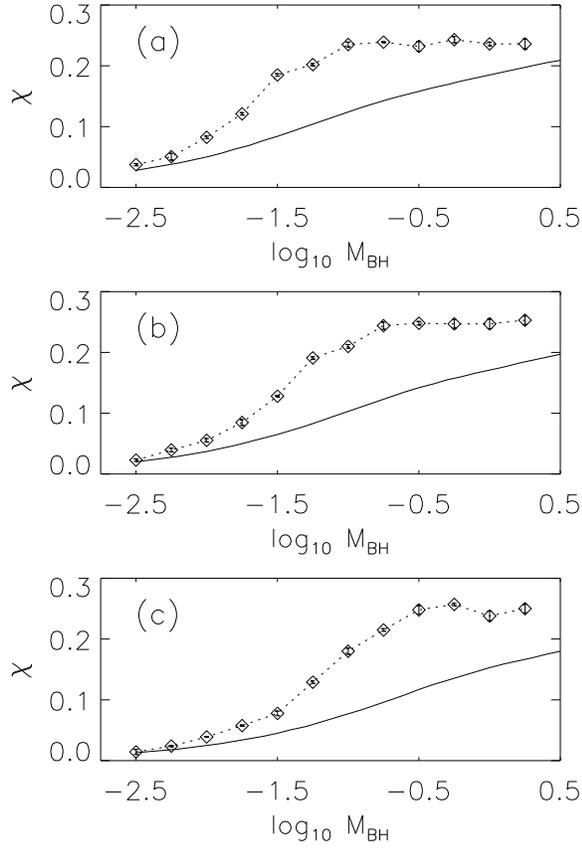}
           }
        \begin{minipage}{10cm}
        \end{minipage}
        \vskip -0.0in\hskip -0.0in
\caption{\label{fig-3}
Same as Fig.~\ref{fig-1}, but with all orbits restricted to the 2-dimensional $(x,y)$-plane.}
\end{figure}

The results of FIGS.~\ref{fig-1} and~\ref{fig-2} suggest that the 
6-dimensional phase space governed by the toy potential~(\ref{eq:toy}) 
exhibits global chaos and associated rapid irreversible mixing over the bulk 
of the parameter space.  Can the same be said for a lower-dimensional analog, 
$\it i.e.$ one corresponding to the toy potential in which $z=p_{z}=0$?  
FIGURE~\ref{fig-3}, which provides the same information as FIG.~\ref{fig-1}, 
but now for a 4-dimensional phase space, hints at the answer.  One now sees 
the agreement between the numerical and analytic results to be less good, as 
would be expected because the fraction of globally chaotic orbits is generally 
much reduced over the 6-dimensional case.  Nonetheless, the results are still 
comparable within a factor of two.

\section{THE SCOPE OF THIS PAPER}
The obvious question is whether the striking agreement between theory
and numerics described in the preceding section is simply a fortuitous 
accident, or whether it is in fact generic. Can the Casetti {\it et al.} 
analysis provide reasonable estimates of the largest Lyapunov exponent for 
generic lower-dimensional Hamiltonian systems? 

Related to this is another important question: To what extent are the 
assumptions implemented by Casetti {\it et al}
justified for lower-dimensional systems? To the extent that they are 
{\it not} justified, one might expect either (i) that the final formula for 
${\chi}$ which they derived is comparatively insensitive to (some of) the 
assumptions and/or (ii) that modifying these assumptions might lead to 
improved estimates.

These questions were addressed by a detailed exploration of orbits in the
potentials discussed in Section II, as well as three other (classes of) 
potentials:
\par\noindent 1. The sixth order truncation of the Toda lattice\cite{To}, 
a familiar two-dimensional potential:
$$V(x,y)={1\over 2}{\Bigl(}x^{2}+y^{2}{\Bigr)}+x^{2}y-{1\over 3}y^{3}
+{1\over 2}x^{4}+x^{2}y^{2}+{1\over 2}y^{4}$$
\begin{equation}
+x^{4}y+{2\over 3}x^{2}y^{3}-{1\over 3}y^{5}
+{1\over 5}x^{6}+x^{4}y^{2}+{1\over 3}x^{2}y^{4}+{11\over 45}y^{6}. 
\end{equation}
\par\noindent 2. A multi-dimensional generalisation of the dihedral
potential\cite{D4}, for one particular set of parameter values, allowing for
$D=2$ through $D=6$:
\begin{equation}
V(q_{1},..,q_{D})=-\sum_{i=1}^{D}q_{i}^{2} +{1\over 4}\left( \sum_{i=1}^{D}
q_{i}^{2}\right)^{2} - {1\over 4} \sum_{i<j=1}^{D}q_{i}^{2}q_{j}^{2}.
\end{equation}
\par\noindent 3. A generalisation of the Fermi-Pasta-Ulam ({\it FPU}) ${\beta}$
model\cite{FPU} with
\begin{equation}
V(q_{1},..,q_{D})=\sum_{i=1}^{D} 
\left[ {a\over 2}\left(q_{i+1}-q_{i})\right)^{2}
+{b\over 4}\left(q_{i+1}-q_{i}\right)^{4} \right],
\end{equation}
with $q_{D+1}{\;}{\equiv}{\;}q_{1}$, allowing for $D=3$ through $D=6$ (the 
special case $D=2$ is integrable). For $a=1$, (3.3) reduces to the standard
{\it FPU} model; for $a<0$, the potential admits extrema that are local maxima,
so that the local mean curvature can become negative. The case $a=1$ and 
$b=0.1$ was considered by Casetti {\it et al.} for much larger values of $D$.

Section IV of this paper begins by providing a terse mathematical summary of
the formalism introduced by Casetti {\it et al.} to estimate the values of the
largest Lyapunov exponent in higher-dimensional systems. This mathematical 
structure is then restated in much simpler physical language and the resulting
reformulation is used to suggest how their analysis could be reformulated for
lower-dimensional systems. Section V summarises the results of extensive
simulations in the potentials (2.1) and (3.1) - (3.3) which were used to test 
the validity of the original assumptions. 
Section VI then turns to the actual values of
Lyapunov exponents estimated using this general approach, considering both the
`true' Lyapunov exponents, defined in a $t\to\infty$ limit, and short-time
Lyapunov exponents\cite{Gr} appropriate for orbit segments of comparatively 
short duration. Estimates of the latter for a variety
of different orbit segments evolved in the same potential with the same energy
reveals an important point: Even when the estimated short-time exponents 
${\chi}_{est}$ differs from the `true' exponents ${\chi}_{num}$ computed 
numerically by as much as a factor of two, their values tend to be strongly 
correlated. For example, orbit segments for which ${\chi}_{est}$ is especially 
small correspond in general to orbits for which the numerical ${\chi}_{num}$ 
is also especially small. In this sense, it is clear that, even if the Casetti
{\it et al.} formula for ${\chi}$ is not completely satisfactory, it 
{\it does} capture some important aspects of the flow. Section VII concludes 
by summarising the principal conclusions and discussing potential implications 
and extensions.
\section{CHAOTIC MOTION AS A STOCHASTIC PARAMETRIC INSTABILITY}
\subsection{ The proposal of Casetti, Clementi, and Pettini} 
The starting point is the reformulation of a
time-independent Hamiltonian system as a geodesic flow in an 
appropriately defined curved space. This can be done in a variety of different
ways, the best known of which involves implementing Maupertuis'
Principle\cite{Lan}, which leads to the Jacobi metric. However, from a 
practical perspective, the most convenient choice is to work with the 
Eisenhart metric \cite{Ei}. 

Given a $D$ degree-of-freedom Hamiltonian system characterised by a Lagrangian
\begin{equation}
L=T-V={1\over 2}a_{ij}{\dot q}^{i}{\dot q}^{j}-V(q^{1}, ... , q^{D}),
\end{equation}
with motion defined on some manifold $M$, consider the extended
manifold $M\times R^{2}$, with coordinates $(q^{0},q^{1},...,q^{D},q^{D+1})$, 
and introduce the Eisenhart metric
\begin{eqnarray}
ds^{2} & = & g_{\mu\nu}dx^{\mu}dx^{\nu} \nonumber\\
       & = & a_{ij}dq^{i}dq^{j}-2V(q)dq^{0}dq^{0} + 2dq^{0}dq^{D+1}. 
\end{eqnarray}
Setting $q^{0}=t$ and $q^{D+1}=t/2-\int_{0}^{t}dt'L({\bf q},{\bf\dot q})$ 
yields $ds^{2}=dt^{2}$. Without loss of generality one can set $a_{ij}=
{\delta}_{ij}$, making the kinetic energy a sum of quadratic contributions
$({\dot q}^{i})^{2}/2$, in which case the geodesic equations reduce to
Newton's equations of motion. Correspondingly the Riemann tensor simplifies
greatly; its only nonvanishing components are
\begin{equation}
R_{0i0j}={{\partial}^{2}V\over {\partial}q^{i}{\partial}q^{j}},
\end{equation}
and the Jacobi equation for a linearised perturbation becomes
\begin{equation}
{\ddot{\xi}}^{i}+{\delta}^{ij}R_{0j0k}{\xi}^{k}=0, 
\qquad (i=1,...,D).
\end{equation}

Were the Riemann components entering into Eq.~(4.4) were everywhere
negative, an arbitrary perturbation would always grow 
exponentially fast. Everywhere negative curvature implies chaotic behavior
and positive Lyapunov exponents\cite{H,A}. The important point, however, is 
that, because of parametric instability, one can have chaotic orbits with 
positive Lyapunov exponents even if the curvature is everywhere positive. If, 
following Casetti {\it et al.}, one assumes that the curvature varies 
`randomly' along a chaotic orbit, eq. (4.5) reduces to a stochastic-oscillator 
equation of the form 
\begin{equation}
{d^{2}{\xi}^{i}\over dt^{2}}
+k^{i}_{\;j}(t){\xi}^{j}=0,
\qquad (i=1,...,D),
\end{equation}
where the matrix $k^{i}_{\;j}$ is characterised completely by its statistical 
properties. However, it is well known that, even if $k^{i}_{\;j}$ is positive 
definite for all times, ${\xi}^{i}$ can grow exponentially\cite{Van}.

Especially in high dimensions, matrix equations become difficult to solve
either numerically or analytically. For this reason, Casetti {\it et al.} 
proceeded by replacing this $D$-dimensional equation with a simpler 
one-dimensional
equation which aims to capture its `average' behavior. Formally, they start
by observing that the Riemann tensor can be decomposed into two pieces:
\begin{equation}
R^{i}_{\;jkl}={1\over D-1}\left( R_{jl}{\delta}^{i}_{k}
 - R_{jk}{\delta}^{i}_{l}\right) + W^{i}_{\;jkl}
\end{equation}
where $R_{ij}$ the Ricci tensor and $W^{i}_{\;jkl}$ is the Weyl projective
tensor. For the case of an isotropic space, the Weyl tensor vanishes 
identically and $R_{jl}{\dot q}^{j}{\dot q}^{l}/(D-1)$
reduces to the (constant) sectional curvature. The crucial assumption then
is that, even though the space is not isotropic, it should appear nearly
isotropic when viewed on comparatively large scales. In the context of such
a `quasi-isotropic' approximation, $R^{i}_{\;k}$ and {\bf I fixed this:} $w^{i}_{\;k}={\delta}^{ij}W^{l}_{\;jlk}$ become diagonal, and Eq~(4.6) reduces to a one-dimensional equation of the form
\begin{equation}
{d^{2}{\xi}\over dt^{2}}+k(t){\xi}=0.
\end{equation}
All that remains is to specify the statistical properties of the random
process $k(t)$.

For a generic Hamiltonian system, the form of $k$ could be quite complex. 
However, given the assumption that $D$ is large, one might
expect that the complicated details will largely wash out. Thus, Casetti 
{\it et al.} argue in the spirit of the Central Limits Theorem that the 
curvature fluctuations in different directions can be approximated as nearly 
independent and, at any instant, Gaussianly distributed. It then follows that
\begin{equation}
{d^{2}{\xi}\over dt^{2}}+{\Omega}{\xi} + {\sigma}{\eta}{\xi}=0,
\end{equation}
where, in terms of the quantity 
$K={\partial}^{2}V/{\partial}q^{i}{\partial}q_{i}{\;}{\equiv}{\;}{\Delta}V$, 
which has mean 
${\langle}K{\rangle}$ and dispersion ${\delta}K$, the quantity
${\Omega}={\langle}K{\rangle}/(D-1)$, ${\sigma}={\delta}K/\sqrt{D-1}$, and
${\eta}$ is Gaussian noise with zero mean and unit variance.
The factors involving $D$ reflect the fact that the curvature-driven
motion in the $i^{th}$ direction is, on the average, `shared' by the 
$D-1$ orthogonal directions. 

Presuming further that the flow is ergodic and that (almost) all orbits
are chaotic, 
the quantities
${\langle}K{\rangle}$ and ${\delta}K$ can be calculated assuming a uniform
sampling of the constant energy hypersurface, {\it i.e.,} a
microcanonical distribution ${\mu}{\;}{\propto}{\;}{\delta}_{D}(H-E)$.
Alternatively, for sufficiently high dimensions one can proceed instead by 
assuming a canonical distribution which, for large $D$, is much simpler 
computationally although it should yield nearly identical results. 

To complete the characterisation of the random process $k(t)$ it remains 
to specify the autocorrelation function ${\Gamma}(t)$ or, at least, the
autocorrelation time ${\tau}$,  which governs how rapidly the curvature
fluctuates along the orbit. This Casetti {\it et al.} provide using
another geometric argument. On the one hand, they identify a time scale
\begin{equation}
{\tau}_{1}{\;}{\approx}{\;}{{\pi}\over 2\sqrt{{\Omega}+{\sigma}}}
\end{equation}
corresponding to the typical time between successive conjugate points, 
{\it i.e.,} points where the Jacobi field of geodesic deviation vanish.
On the other, they identify a time scale
\begin{equation}
{\tau}_{2}{\;}{\approx}{\;}{{\Omega}^{1/2}\over {\sigma}}
\end{equation}
corresponding to the length scale on which the fluctuations become comparable
to the average curvature. They then select as an appropriate autocorrelation
time a value ${\tau}$ satisfying
\begin{equation}
{\tau}^{-1}=2\left( {\tau}_{1}^{-1} + {\tau}_{2}^{-1} \right)
\end{equation}
so that
\begin{equation}
2{\tau}={{\pi}\sqrt{{\Omega}}\over 
2\sqrt{ {\Omega} \left( {\Omega}+{\sigma} \right)}
+ {\pi}{\sigma}}
\end{equation}
Casetti {\it et al.} suggest further that ${\Gamma}(t)$
might be well approximated by the oscillating function
\begin{equation}
{\Gamma}(t)={{\Omega}^{2}\over {\pi}}\,{\sin {\omega}t \over {\omega}t},
\end{equation}
which yields an autocorrelation time
\begin{equation}
{\tau}={1\over 2{\omega}}={{\tau}_{osc}\over 4{\pi}},
\end{equation}
with ${\tau}_{osc}$ the oscillation period.
However, this is largely irrelevant for their analysis. Granted the assumption 
of additive Gaussian noise, the form of the color only enters into the final expression for ${\chi}$ through the autocorrelation time ${\tau}$\cite{Van}.

Given a knowledge of ${\tau}$ and the first two moments, Eq. (4.8) can be
solved analytically using a technique developed by van Kampen\cite{Van}
to yield an estimated largest Lyapunov exponent
\begin{equation}
{\chi}={1\over 2}\left( {\Lambda}-{4{\Omega}\over 3{\Lambda}}\right),
\end{equation}
where
\begin{equation}
{\Lambda}=\left[ 2{\sigma}^{2}{\tau} + 
\sqrt{ \left( {4{\Omega}\over 3} \right)^{3} +
\left( 2{\sigma}^{2}{\tau}\right)^{2}}\; \right]^{1/3}.
\end{equation}

\subsection{ Applying this proposal to lower-dimensional Hamiltonian systems}
The preceding can be reformulated without recourse to differential 
geometry in a setting that makes the physical content of the assumptions more
transparent and, as such, makes it clearer which assumptions might prove 
suspect, especially for lower-dimensional systems.

The basic perturbation equation (4.5) can be derived trivially from the 
original Hamilton equations
\begin{equation}
{dq^{i}\over dt}={{\partial}H\over {\partial}p_{i}}={\delta}^{ij}p_{j}
\end{equation}
and
\begin{equation}
{dp_{i}\over dt}=-{{\partial}H\over {\partial}q^{i}}
=-{{\partial}V\over {\partial}q^{i}}
\end{equation}
associated with the Hamiltonian
\begin{equation}
H=T+V={1\over 2}{\delta}^{ij}p_{i}p_{j}+V(q).
\end{equation}
It is clear that the introduction of a small perturbation
$q^{i} \to q^{i}+{\xi}^{i}$ and $p_{i}\to p_{i}+{\zeta}_{i}$ leads to evolution equations of the form
\begin{equation}
{d{\xi}^{i}\over dt}={\delta}^{ij}{\zeta}_{j} \qquad {\rm and} \qquad
{d{\zeta}_{i}\over dt}=-{{\partial}^{2}V\over {\partial}q^{i}{\partial}q^{j}}
{\xi}^{j},
\end{equation}
but combining these last two equations leads immediately to Eq.~(4.4).

The crucial assumption underlying the entire Casetti {\it et al.} analysis is 
the assumption that, for the case of chaotic orbits, Eq.~(4.4) can be modelled
by a stochastic-oscillator equation. For the case of `wildly chaotic' orbits
or orbit segments, which are far from periodic, this assumption would seem
quite reasonable. However, in lower dimensional systems one encounters the 
possibility of `sticky'\cite{GC} orbit segments which, albeit characterised
by positive short-time Lyapunov exponents, are `nearly regular' in visual
appearance and have Fourier spectra with most of the power concentrated at or
near a few special frequencies\cite{PK}. This is especially common for $D=2$, 
where cantori\cite{LL} can serve as entropy barriers, confining a chaotic 
orbit near a regular island for surprisingly long times. To the extent that 
such orbit segments behave in a nearly regular fashion, the assumption of nearly
random behavior is clearly suspect, and one might anticipate that a stochastic
oscillator equation cannot prove completely satisfactory. Alternatively, to the
extent that this `sticky' behavior is rare, such an equation might be expected to provide a reasonable starting point.

The assumption of `quasi-isotropy' can also be understood in very simple 
physical terms: Instead of considering the $D$-dimensional Eq.~(4.4), 
which involves the full second derivative matrix of $V$, it is assumed that, 
on the average, each direction in configuration space is statistically identical, so that one can consider instead $D$ identical one-dimensional equations. In this context, the only question concerns the proper identification of the quantity to play the role of the squared frequency $k(t)$. The Casetti {\it et al.} prescription atates that the relevant information about stability is contained in the trace of the second derivative matrix, so that $k(t)$ should
be proportional to ${\Delta}V={\partial}^{2}V/{\partial}q^{i}{\partial}q_{i}$.
The factor of $D-1$ entering into Eq.~(4.8) reflect the fact that the 
perturbation driving the chaos is `shared' amongst $D-1$ dimensions. (Recall
that, in a time-independent Hamiltonian system, there is always one direction
of neutral stability corresponding to translation along the orbit from 
$q^{i}(t)$ to $q^{i}(t+{\delta}t)$.)

This assumption of quasi-isotropy seems especially reasonable for large $D$
where, on average, different directions of the configuration space should
look much the same, but becomes more suspect in lower dimensions. In principle,
one can relax this assumption by working with the full matrix equation. As
a practical matter, however, this becomes quite cumbersome for $D{\;}{\gg}{\;}
2$. For this reason, most the following analysis will retain the assumption of
quasi-isotropy. What happens when this assumption is relaxed for 
two-dimensional systems is considered briefly in Section VI. It would in fact
appear that, at least for $D=2$, relaxing this assumption does not in general
yield significant improvement in the estimated value of the largest Lyapunov
exponent.

For generic Hamiltonian systems with large $D$ one anticipates that (almost) 
all the orbits on a constant-energy hypersurface are chaotic. Granted the
assumption of ergodicity, it then follows that, over sufficiently long time
scales, an orbit eventually samples a microcanonical distribution. This implies
that, when estimating a Lyapunov exponent ${\chi}(E)$, as defined in an
asymptotic $t\to\infty$ limit, one can assume that the 
statistical properties of the curvature experienced by an orbit are given 
correctly by a microcanonical distribution. By contrast, for lower dimensional 
systems, especially $D=2$, one anticipates instead that a generic potential
will admit a coexistence of large measures of both regular and chaotic orbits,
so that the assumption of a microcanonical distribution is {\it not}
justified. Rather, granted the assumption of ergodicity, one would anticipate
the existence of a different invariant distribution, corresponding to a 
uniform population of those portions of the constant energy hypersurface 
which are accessible to a chaotic orbit with specified initial condition. 
It is not clear how this distribution could be computed analytically. However,
as described in Section V, numerical approximations to this invariant 
distribution can be generated straightforwardly through a time-series analysis 
of orbits evolved numerically. 

Even if a microcanical distribution is justified, the assumption of Gaussian 
fluctuations is problematic. For large $D$, this assumption can be motivated 
with a fair degree of rigor via a Central Limits Theorem argument, supposing 
that the distribution of values of ${\Delta}V$ involves a convolution of 
$D$ nearly independent distributions for the separate components 
${\partial}^{2}V/{\partial}q^{i}{\partial}q^{i}$ (no sum over indices).
For very small $D$, this is clearly not justified, and the minimum value of
$D$ for which the Gaussian approximation {\it is} justified must depend to
a certain extent on the form of the individual distributions. It will be seen 
in Section V that, for the model systems (3.2) and (3.3), the convergence 
towards a Gaussian is quite efficient, and that the distribution 
$N[{\Delta}V]$ is reasonably well fit by a Gaussian even for $D$ as small as 
$3$ or $4$. It will also be seen that, at least for distributions 
$N[{\Delta}V]$ that are not too skew, deviations from a Gaussian do not change 
the estimated value of the largest Lyapunov exponent all that much.

The formula for the autocorrelation time ${\tau}$ motivated by Casetti {\it 
et al.} is somewhat {\it ad hoc} in that, unlike the other crucial
inputs ${\Omega}$ and ${\sigma}$,  it cannot be derived directly from a 
microcanonical distribution. However, the basic scaling implicit in ${\tau}$
can again be inferred relatively simply. As will be seen below, ${\Omega}$ and 
${\sigma}$ are typically comparable in magnitude. They both reflect 
statistical properties of ${\Delta}V$ and, as such scale (within factors of
order unity) as ${\overline V}/R^{2}$, where ${\overline V}$ represents a
typical value of potential and $R$ is a characteristic length scale, 
{\it i.e.,} the size of the configuration space region probed by an orbit. 
Assuming `virialisation,' {\it i.e.}, that the mean potential and kinetic 
energies of the orbits are comparable in magnitude, it follows that 
${\overline V}{\;}{\sim}{\;}v^{2}$, where $v$ denotes a typical orbital speed.
However, this implies that
${\Omega}{\;}{\sim}{\;}{\sigma}{\;}{\sim}{\;}v/R{\;}{\equiv}{\;}t_{D}^{-1}$, 
where
$t_{D}$ denotes a characteristic dynamical or crossing, time. Allowing for
the fact that the characteristic scale on which $V$ changes significantly is 
typically somewhat smaller than the size of region accessible to the orbit
leads to the obvious physical conclusion that ${\tau}$ should be comparable to,
but somewhat smaller than, the time required for an orbit to travel from one
side of the system to another. 

Implicit in the Casetti {\it et al.} analysis is the assumption that the 
stochastic process $k(t)$ corresponds to state-independent, additive noise, 
so that, {\it e.g.,} the autocorrelation time ${\tau}$ on which the curvature
changes is independent of the value of the curvature.
Strictly speaking, this assumption cannot be correct.
If, {\it e.g.,} an orbit is in a region where $V$ is especially small, its
kinetic energy, and hence its velocity, will be especially large, so that
the orbit will move very quickly to a different region where $V$, and hence in 
general ${\Delta}V$, is very different. If, instead, the orbit is in a region
where $V$ is especially large, it will move more slowly so that ${\Delta}V$
might be expected to change more slowly. The autocorrelation time ${\tau}$ of
(4.13) represents an average over a variety of orbits with very different
values of $V$. One might anticipate that these differences will tend to wash
out for large $D$, but there is no obvious reason why this should be true for
smaller $D$. 

One final point: It is clear that, for small $D$, one cannot pass from a
microcanonical to a canonical description. One must work directly with the
microcanonical measure ${\mu}{\;}{\propto}{\;}{\delta}_{D}(H-E)$. This, 
however, is not a major problem. For a $D$ degree of freedom system, the
microcanonical distribution corresponds to a configuration space density
\begin{equation}
f(q^{i}){\;}{\propto}{\;}\cases{ \left( E-V \right)^{(D-2)/2} & if
$V{\;}{\le}{\;}E$;\cr 
 & \cr
0 & if $V>E$.\cr}  
\end{equation}
but, given this formula for $f$, it is straightforward, at least numerically,
to compute 
the distribution $N[{\Delta}V]$ and/or any moments of the distribution.

\section{TESTING THE BASIC ASSUMPTIONS}
\subsection{What was computed}
To test the validity of the basic assumptions requires a comparison of real 
orbital data with predictions made assuming a microcanonical distribution. 
The requisite orbital data were generated and analyzed as follows: 

For given choices of potential and total energy, a collection of $N=1000$ 
initial conditions were selected, and each of these was integrated into the future for a total time $T$ corresponding to between ${\sim}{\;}100$ and $2000$ characteristic crossing times $t_{D}$. The numerical integration simultaneously tracked the evolution of a small initial perturbation, periodically renormalised in the usual way\cite{LL} so as to yield an estimate of the largest
(short-time) Lyapunov exponent for the orbit segment. Configuration space 
data, recorded at fixed intervals ${\delta}t$, were used to generate a time 
series $\{K_{j}(n{\delta}t)\}$ for each of the segments in the $1000$ orbit 
ensemble which was deemed to be chaotic. ${\delta}t$ was typically so chosen
that each segment was sampled by $2560$ points. Distinctions between regular 
and chaotic were implemented through the introduction of a threshhold value 
${\chi}_{min}$: if the computed ${\chi}<{\chi}_{min}$, the orbit segments were 
assumed to be regular. Combining all the orbital data for all the chaotic 
orbits allowed the computation of the bulk moments ${\langle}K{\rangle}$ and 
${\delta}K$ where, recall, $K={\Delta}V$. Binning the combined data into 1000 
bins yielded a numerical representation of the distribution $N[K]$. 

A discretized representation of the average autocorrelation function 
${\Gamma}(t)$ was computed by selecting a representative ensemble of $5120$ 
initial conditions, evolving each of these into the future for an extended 
time $T{\;}{\ge}{\;}2048$, so as to generate a set of well-mixed `random' 
phase space points, identifying each of the $N_{c}{\;}{\le}{\;}N$ orbit 
segments that were chaotic, and, by extending the integrations for an 
additional time $T'=1024$, constructing
\begin{equation}
{\Gamma}(n{\delta}t)={1\over N_{c}{\langle}K^{2}{\rangle}}\;
\sum_{j=1}^{N_{c}}\;DK_{j}(T)
DK_{j}(T+n{\delta}t).
\end{equation}
Here $DK_{j}{\;}{\equiv}{\;}K_{j}-{\langle}K{\rangle}$, and the quantities 
${\langle}K{\rangle}$ and ${\langle}K^{2}{\rangle}$ represent averages computed
for all the chaotic orbital data for $T<t<T'$. Ideally one should compute the 
autocorrelation time ${\tau}$ using the defining relation
\begin{equation}
\int_{0}^{\infty}\,dt\,{\Gamma}(t) = {\langle}K^{2}{\rangle}{\tau}.
\end{equation}
Given, however, that ${\Gamma}$ is typically a rapidly oscillating function 
(period ${\sim}{\;}t_{D}$) with an envelope that damps very slowly, such a computation proved unreliable. A seemingly better measure of ${\tau}$ or, at least, of how ${\tau}$ scaled with energy $E$ for fixed potential, was obtained by computing the period ${\tau}_{osc}$ associated with the oscillations. 

Predictions associated with a microcanonical distribution were computed as
follows: The microcanonical distribution 
${\mu}{\;}{\propto}{\;}{\delta}_{D}(H-E)$ implies the configuration space
probability density (4.22); but, given this $f$, it is straightforward to
compute the value of any configuration space function $g(q)$.
Numerical representations of the distribution $N[K]$ associated with a 
microcanonical distribution were computed 
by dividing the occupied configuration space into a collection
of $M$ hypercubes, (ii) deciding randomly whether or not to sample each
hypercube, using a weighting ${\propto}{\;}(E-V)^{(D-2)/2}$ as evaluated
at a random point in the cube, (iii) in the event that the hypercube was to be 
sampled, locating a point in the cube at a randomly chosen location, and then
(iv) binning the resulting collection of points into 1000 bins.

Granted the assumption of a Gaussian distribution of curvatures, estimates of
the Lyapunov exponent ${\chi}$ can be, and were, computed using eq.~(4.16),
which does not require the assumption of a microcanonical population. When
the assumption of a Gaussian distribution is relaxed, an analytic solution 
is not possible in general, so that ${\chi}$ was obtained instead from a
numerical computation, with the random curvature generated initially by
sampling $N[K]$, held constant for the autocorrelation time ${\tau}$, and then 
replaced by another, randomly chosen curvature\cite{Kan2}.
\subsection{What was found}
\subsubsection{$N[K]$ and its first two moments}

\begin{figure}[t]
\centering
\centerline{
        \epsfxsize=8cm
        \epsffile{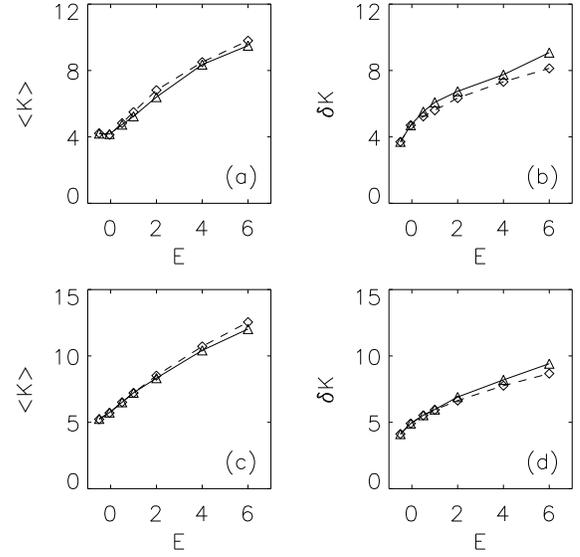}
           }
        \begin{minipage}{10cm}
        \end{minipage}
        \vskip -0.0in\hskip -0.0in
\caption{\label{fig-4}
(a) The mean curvature ${\langle}K{\rangle}$ for chaotic orbits
in the $D=2$ dihedral potential as a function of energy $E$, computed assuming
a microcanonical distribution (solid line) and extracted directly from orbital
data (dashed line). (b) The associated dispersion ${\delta}{K}$. (c) 
${\langle}K{\rangle}$ for $D=3$. (d) ${\delta}{K}$ for $D=3$.}
\vspace{-0.0cm}
\end{figure}

FIGURE~\ref{fig-4} exhibits the energy-dependence of the quantities 
${\langle}K{\rangle}$ and ${\delta}K$ for chaotic orbits in the dihedral
potential with $D=2$ and $D=3$, computed both from time-series data
(dashed lines) and assuming a microcanonical distribution (solid lines).
Overall, one observes excellent agreement between the numerical and analytic
predictions, particularly for the first moment ${\langle}K{\rangle}$.
The best overall agreement obtains
for lower energies where, even for $D=2$, the measure of regular orbits is
comparatively small and `stickiness' seems comparatively unimportant. 

\begin{figure}[t]
\centering
\centerline{
        \epsfxsize=8cm
        \epsffile{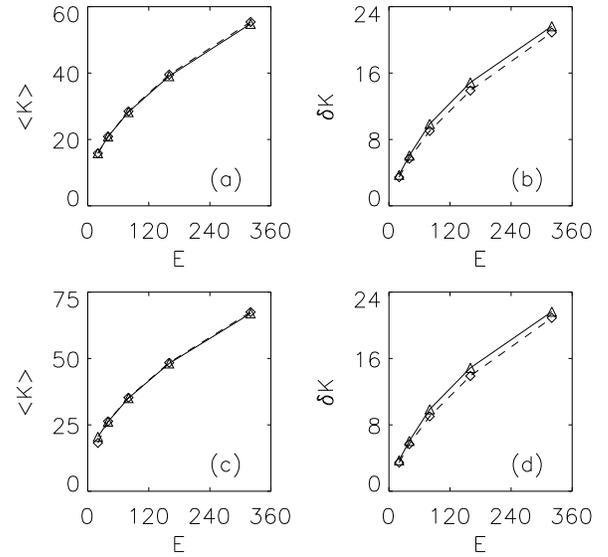}
           }
        \begin{minipage}{10cm}
        \end{minipage}
        \vskip -0.0in\hskip -0.0in
\caption{\label{fig-5}
(a) The mean curvature ${\langle}K{\rangle}$ for chaotic orbits
in the $D=4$ {\it FPU} potential with $a=1.0$ and $b=0.1$ as a function of 
energy $E$, computed assuming
a microcanonical distribution (solid line) and extracted directly from orbital
data (dashed line). (b) The associated dispersion ${\delta}{K}$. (c) 
${\langle}K{\rangle}$ for $D=6$. (d) ${\delta}{K}$ for $D=6$.}
\vspace{-0.0cm}
\end{figure}

For $D=2$ at higher energies, say $E>1.0$ or so, it appears that a third of 
the constant energy hypersurface, or even more, corresponds to regular orbits, 
so that one is clearly {\it not} justified in assuming a 
microcanonical distribution. However, it is evident that the predictions based 
on a microcanonical distribution remain quite good. That this should be the 
case is not really surprising. Presuming that the regular islands are not 
concentrated preferentially at regions where ${\Delta}V$ is especially large 
or small, it would seem reasonable to assume that, in a sufficiently  
coarse-grained 
sense, chaotic orbits still go `all over' the energetically allowed regions of
configuration space. To the extent, however, that this be true, one might
expect moments approximating the moments appropriate for a microcanonical
distribution which, for $D=2$, implies [cf.~Eq. (4.21)] a uniform 
configuration space density.
FIGURE~\ref{fig-5} exhibits analogous data for the {\it FPU} potential with 
$D=4$ and $D=6$, generated for parameter values $a=1.0$ and $b=0.1$.

\begin{figure}[t]
\centering
\centerline{
        \epsfxsize=8cm
        \epsffile{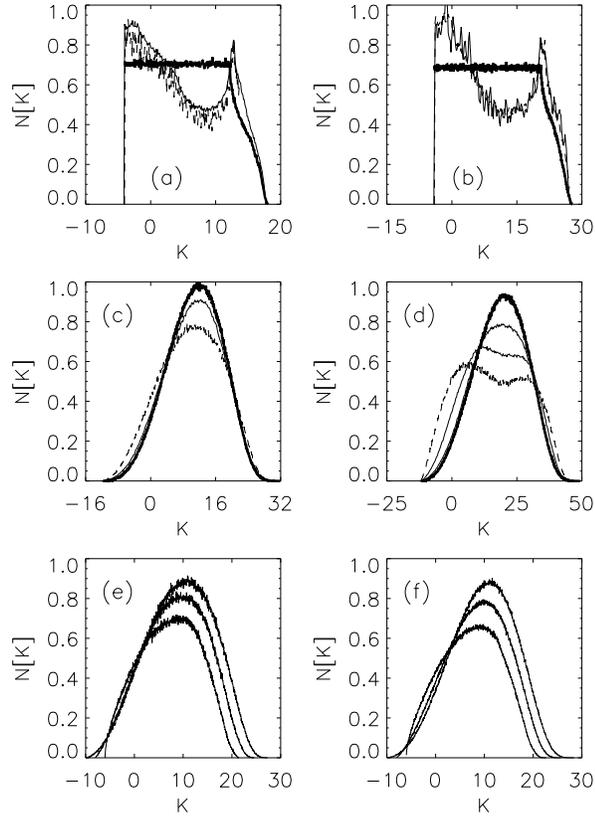}
           }
        \begin{minipage}{10cm}
        \end{minipage}
        \vskip -0.0in\hskip -0.0in
\caption{\label{fig-6}
(a) The distribution of curvatures, $N[K]$, for chaotic orbits with
energy $E=1.0$ in the $D=2$ dihedral potential, computed assuming a 
microcanonical distribution (thick solid line) and from orbital data for an
ensemble evolved for times $t=1024$ and $t=3172$. (b) 
$N[K]$ for $D=2$ and $E=6.0$. 
(c) $N[K]$ for $D=6$ and $E=1.0$. (d) $N[K]$ for $D=6$ and $E=6.0$.
(e) $N[K]$, as generated from time-series data for $t=3172$, for $E=1.0$ with 
$D=3$, $D=4$, and $D=5$. (f) $N[K]$ for $E=1.0$ with $D=3$, $D=4$, and $D=5$, 
now generated assuming a microcanonical distribution.
}
\vspace{-0.0cm}
\end{figure}

\begin{figure}[t]
\centering
\centerline{
        \epsfxsize=8cm
        \epsffile{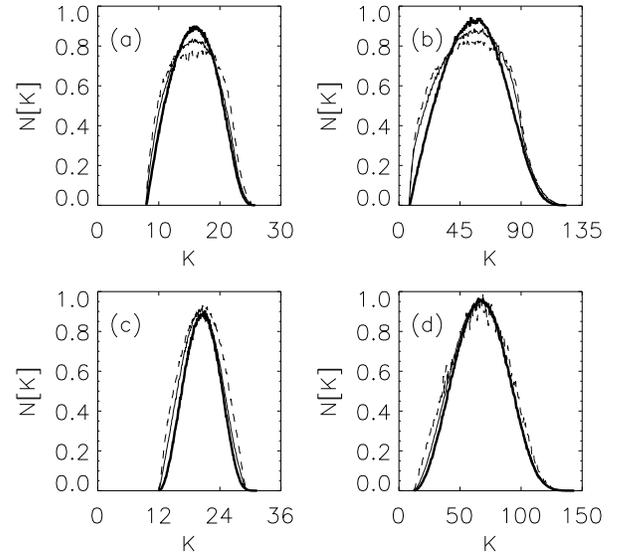}
           }
        \begin{minipage}{10cm}
        \end{minipage}
        \vskip -0.0in\hskip -0.0in
\caption{\label{fig-7}
(a) The distribution of curvatures, $N[K]$, for chaotic orbits with
energy $E=20$ in the $D=4$ {\it FPU} potential with $a=1.0$ and $b=0.1$, 
computed assuming a 
microcanonical distribution (thick solid line) and from orbital data for an
ensemble evolved for times $t=1024$ and $t=4196$. (b) 
$N[K]$ for $D=4$ and $E=320$. 
(c) $N[K]$ for $D=6$ and $E=20$. (d) $N[K]$ for $D=6$ and $E=320$.
}
\vspace{-0.0cm}
\end{figure}

The thick solid curves in panels (a) - (d) of FIG.~\ref{fig-6} exhibit 
distributions 
of curvatures, $N[K]$, for the dihedral potential with $D=2$ and $D=6$ 
generated assuming a microcanonical distribution. The corresponding curves in 
FIG~\ref{fig-7} exhibit analogous distributions for the {\it FPU} potential 
for $D=4$ and $D=6$. Panels (e) and (f) in FIG.~\ref{fig-6} show the time 
series and microcanonical predictions
for the dihedral potential in (from left to right) $D=3$, $4$, and $5$.
It is evident that the microcanonical distributions for the dihedral potential
with $D=2$ are not even remotely Gaussian in shape. However, it is also 
apparent that, for all the other cases, the distribution is in fact reasonably
well fit by a Gaussian, although $N[K]$ typically has a slight skew and can
manifest appreciable deviations for large $|K-{\langle}K{\rangle}|$.

The other curves in FIG.~\ref{fig-6} (a) - (d) and in FIG.~\ref{fig-7} 
represent 
distributions $N[K]$ generated from time-series data. FIG.~\ref{fig-7} and the 
first three panels of FIG.~\ref{fig-6} display two numerical curves, one 
representing data for $0<t<1024$ and the other for $2048<t<3072$. 
FIG.~\ref{fig-6} (d) also includes a third numerical 
curve, generated for $8192<t<9216$. For the two energies exhibited in the 
$D=2$ dihedral potential, $E=1.0$ and $E=6.0$, there exist large measures of 
both regular and chaotic orbits and, for this reason, the time-series $N[K]$ 
differs significantly from the microcanonical $N[K]$. However,
the ensembles of initial conditions used to generate the time-series
distributions evolved towards an invariant (albeit non-microcanonical)
distribution relatively quickly, so that the two numerical curves very nearly
overlap. 

For the dihedral potential with $D{\;}{\ge}{\;}3$ and for the {\it FPU} 
potential with $D{\;}{\ge}{\;}4$ almost all the orbits appear to be chaotic, 
so that, assuming ergodicity, the microcanonical $N[K]$ and a truly 
representative time-series $N[K]$ should coincide up to statistical
uncertainties. However, for the cases exhibited in FIGS.~\ref{fig-6} (c) and 
(d) and FIG.~\ref{fig-7}, the
initial ensembles only converged towards an invariant distribution on a
comparatively long time scale, so that the two (or more) time-series curves
differ appreciably from one another. In each case, the later time sampling(s)
yielded distributions $N[K]$ that more closely approximated the microcanonical
$N[K]$. 

The preceding suggests that one can use the form of the distribution $N[K]$
as a robust diagnostic in terms of which to probe the approach towards
ergodicity. Ergodicity {\it per se} is an assumption regarding the $t\to\infty$
limit and, even assuming ergodicity, there remains an obvious
question: How long must one evolve some ensemble of initial conditions before
its time-averaged density closely approximates the density associated with
a constant population of the accessible phase space regions? Comparing the
distribution $N[K]$ associated with an evolving ensemble with the $N[K]$
associated with a microcanonical distribution can provide a useful diagnostic
for probing the extent to which the ensemble has evolved 
towards a time-independent invariant distribution.

It is well known that different chaotic orbit segments in the same connected
phase space region can exhibit vastly different short-time Lyapunov exponents,
and that the values of these short-time exponents can correlate significantly
with position. For example, chaotic segments near regular islands tend to
be much less unstable than wildly chaotic segments located in the middle of
the stochastic sea. One might, therefore, expect that orbit segments with
especially large or small short-time exponents would be characterised by 
different curvatures. For potentials and energies where almost all the
orbits are chaotic and `stickiness' is rare, this segregation should be 
minimal; but for potentials where there is a coexistence of large measures
of both regular and chaotic orbits, and where `stickiness' is pronounced,
this effect should be much more pronounced.

\begin{figure}[t]
\centering
\centerline{
        \epsfxsize=8cm
        \epsffile{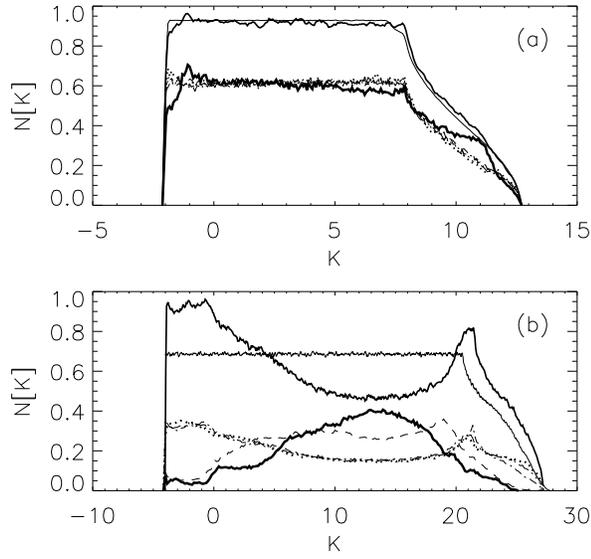}
           }
        \begin{minipage}{10cm}
        \end{minipage}
        \vskip -0.0in\hskip -0.0in
\caption{\label{fig-8}
(a) The distribution of curvatures, $N[K]$, for a representative
ensemble of $1000$ orbits with energy $E=-0.5$ in the $D=2$ dihedral potential.
The five lower curves represent subdistributions, generated by dividing the 
ensemble into five quintiles based on the values of the short-time Lyapunov
exponents for the orbits. The near-horizontal upper curve represents the
distribution $N[K]$ predicted by a microcanonical distribution; the other, more
jagged upper curve represents the distribution $N[K]$ associated with the
full 1000 orbit ensemble, given by the sum of the five lower curves. (b) 
The same for $E=6.0$.}
\vspace{-0.0cm}
\end{figure}

As illustrated in FIG.~\ref{fig-8}, this intuition was corroborated 
numerically. The top panel of FIG.~\ref{fig-8} was generated for $E=-0.5$ in 
the $D=2$ dihedral
potential, an energy where the regular regions are extremely small, so that a
representative ensemble of $1000$ initial conditions, integrated for a time
$T=1024$, yielded no regular orbits. The orbits generated from these initial
conditions were divided into five quintiles, depending on the values of their
short-time Lyapunov exponents, and the lower five curves in this panel 
exhibit individual subdistributions $N[K]$ computed for each quintile. The 
four quintiles corresponding to the larger values of ${\chi}$ yielded
distributions that were nearly identical. The lowest quintile was
again quite similar, but did manifest some noticeable differences: This
subdistribution, corresponding to the thick solid line, is distinctly 
underrepresented at very low values of $K$ and over-represented at very 
large $K$, and, unlike the other four quintiles, appears to be a slowly
decreasing function of $K$ in the interval $0<K<7.5$. The sum of these five
subdistributions (with a different normalisation from the subdistributions) 
corresponds to the slightly 
jagged upper curve, which is essentially identical, modulo statistical 
uncertainties, to the smoother curve computed for a microcanonical 
distribution. 

The lower panel of FIG~\ref{fig-8} exhibits analogous data for $E=6.0$, again 
in the $D=2$ dihedral potential. In this case, a $1000$ orbit ensemble was 
divided instead into a `quintile' of $332$ regular orbits and four `quintiles' 
each comprised of $167$ chaotic orbits, but the resulting subensembles were
analyzed identically. The lower solid curve peaking at $K{\;}{\sim}{\;}13$
represents the $332$ regular orbits, and the three nearly identical curves
which have a local minimum  at $K{\;}{\sim}{\;}13$ correspond to the chaotic
orbit segments with the largest short-time Lyapunov exponents. The intermediate
dashed curve corresponds to the chaotic orbits with the smallest short-time
Lyapunov exponent, the majority of which could be reasonably classified as
`sticky.' The total $N[K]$ given as a sum of the four chaotic `quintiles' is
represented by the upper curve with a local minimum at $K{\;}{\sim}{\;}13$. 
The upper curve corresponding to a nearly flat profile again corresponds to a 
microcanonical distribution.

\subsubsection{The autocorrelation time ${\tau}$}
\begin{figure}[t]
\centering
\centerline{
        \epsfxsize=8cm
        \epsffile{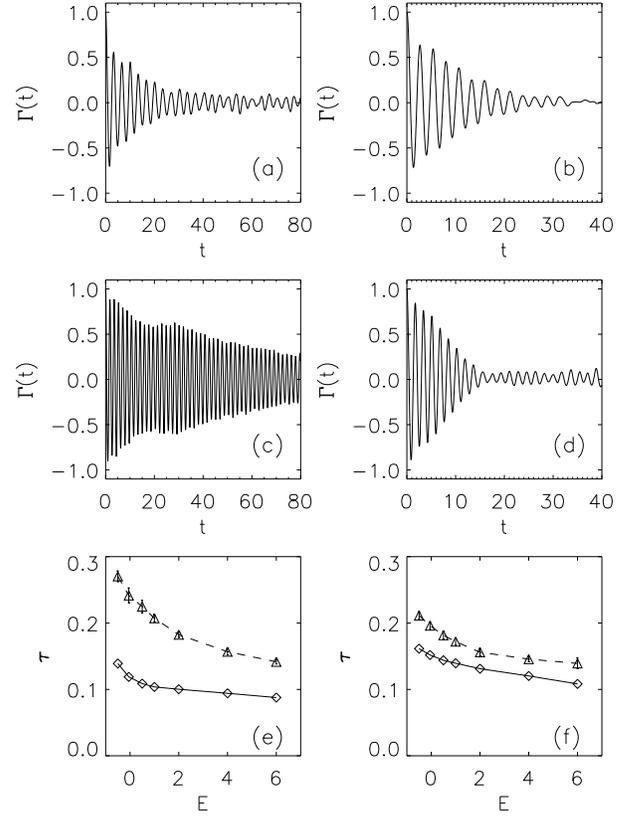}
           }
        \begin{minipage}{10cm}
        \end{minipage}
        \vskip -0.0in\hskip -0.0in
\caption{\label{fig-9}
(a) The autocorrelation function ${\Gamma}(t)$ for chaotic orbits
in the $D=2$ dihedral potential with $E=-0.5$. 
(b) The same for $D=6$ and $E=-0.5$. 
(c) The same for $D=2$ and $E=6.0$. 
(c) The same for $D=6$ and $E=6.0$. 
(e) The Casetti{\it et al.} time scale ${\tau}$ of Eq.~(4.12) (solid line) and 
the time scale ${\tau}_{osc}/4{\pi}$ of Eq.~(4.14) (dashed line), 
for chaotic orbits in the $D=2$ dihedral 
potential at different energies $E$. (f) The same for $D=6$.}
\vspace{-0.0cm}
\end{figure}

As suggested by Casetti {\it et al.}, the autocorrleation function 
${\Gamma}(t)$
is in fact an oscillating function of time, but it tends to decay more slowly 
than with the $1/t$ envelope implicit in Eq.~(4.14). This slower decay is 
especially evident for potentials and energies when `stickiness' is important,
in which case a substantial `memory' can persist for dozens of oscillations.
This is, {\it e.g.,} evident in FIG.~\ref{fig-9} (a) - (d), which exhibit data 
for the dihedral potential for $D=2$ and $D=6$. The first two panels correspond
to a very low energy $E=-0.5$, where, even for $D=2$, almost all the orbits
are chaotic. The second two panels correspond to a higher energy $E=6.0$
where, for both $D=2$ and $D=6$, chaotic segments can be nearly periodic and
have comparatively small short-time Lyapunov exponents. For $D=2$, the case
exhibited in panel (c), roughly one quarter of the chaotic orbits are
noticeably `sticky'; for $D=6$, the case in panel (d), the fraction is
reduced to about 5\%. In either case, analysis of a sample that excludes 
`sticky' segments yields an autocorrelation function that decays substantially 
more quickly. 

As noted by Casetti {\it et al.}, if the autocorrelation function is in fact
well approximated by Eq.~(4.13) the time scale identified geometrically in
Eq.~(4.12) should coincide with the time scale (4.14). This prediction was
tested numerically and found typically to be satisfied to within factors
of ${\sim}$2, although some discrepancies were observed. As noted already, a 
direct determination of ${\tau}$ using Eq.~(5.2) proved unreliable.

Perhaps the most striking point is that, at least when `stickiness' is
comparatively unimportant, the Casetti {\it et al.} time scale ${\tau}$ given
by Eq.~(4.13) and the time scale ${\tau}_{osc}$ exhibit very similar scaling
with energy $E$. This is illustrated in FIG.~\ref{fig-9} (e) and (f), which 
exhibit 
both time scales as functions of $E$ for the $D=2$ and $D=6$ dihedral 
potential. In each case, the time scale ${\tau}_{osc}$ is somewhat longer than 
the Casetti {\it et al.} time scale ${\tau}$. Significantly, though, for
$D=2$ the quantities ${\tau}_{osc}$ and ${\tau}$ exhibit very different 
scalings at higher energies, precisely where `stickiness' is most prominent.

\subsubsection{Sources of uncertainty}
Granted the assumption of a Gaussian distribution of curvatures, the predicted
value of the largest Lyapunov exponent depends on only three quantities, namely
${\langle}K{\rangle}$, ${\delta}K$, and ${\tau}$; and as such, it is natural
to ask how the predicted value ${\chi}_{est}$ varies if any of these inputs
are changed.

If one introduces a simulataneous scaling of both ${\langle}K{\rangle}$ and 
${\delta}K$,
{\it i.e.,} ${\langle}K{\rangle}\to{\alpha}{\langle}K{\rangle}$ {\it and}
${\delta}K\to{\alpha}{\delta}K$, with ${\alpha}$ of order unity, ${\chi}_{est}
\to{\alpha}^{1/2}{\chi}_{est}$. If, alternatively,${\langle}K{\rangle}$
is held fixed but ${\delta}K\to{\alpha}{\delta}K$, one infers, at least 
approximately, that ${\chi}_{est}\to{\alpha}^{3/2}{\chi}_{est}$. Finally,
if ${\langle}K{\rangle}$ and ${\delta}K$ are held fixed, but one allows for
a scaling ${\tau}\to{\alpha}{\tau}$, ${\chi}_{est}\to{\alpha}{\chi}_{est}$.

Given that the values of ${\langle}K{\rangle}$ and ${\delta}K$ can be estimated
quite well using simple dimensional arguments -- recall that, even when there
are relatively large measures of periodic orbits, a microcanonical population
yields estimates in good agreement with the results of direct numerical 
computation --, it would seem that, with the assumption of quasi-isotropy and 
a Gaussian distribution of 
curvatures, the largest source of error should be in the determination of the
autocorrelation time ${\tau}$. The expression for ${\tau}$ motivated by Casetti
{\it et al.} is more {\it ad hoc} than the expression for ${\langle}K{\rangle}$
and ${\delta}K$; and dimensional arguments are hard pressed to yield estimates
of ${\tau}$ that are accurate to better than a factor of two. However, factors
of two uncertainty in ${\tau}$ translate directly into factors of two 
uncertainty in ${\chi}_{est}$.

One can also investigate how the predicted ${\chi}_{est}$ changes if one
relaxes the assumption of a Gaussian distribution, instead computing 
${\chi}_{est}$ by solving Eq.~(4.9) numerically for the distribution $N[K]$ 
generated either from a microcanonical distribution or from real orbital
data. The resulting change in ${\chi}_{est}$ will of course depend on the
degree to which $N[K]$ deviates from a Gaussian, larger deviations resulting
in larger changes. Especially for two-dimensional systems, whre $N[K]$ is far
from Gaussian, allowing for the correct distribution can change ${\chi}_{est}$
by factor of three, or even more.

This is illustrated in FIG.~\ref{fig-9}, which exhibits several different 
estimates of the largest Lyapunov exponents ${\chi}_{est}$ for the $D=2$ 
dihedral potential, most of which will be described in Section VI. In the
present context, note simply (i) the `true' ${\chi}_{num}$, generated by
tracking a small initial perturbation (solid line), (ii) the Casetti {\it 
et al.} ${\chi}_{est}$, based on the assumption of Gaussian fluctuations and
an autocorrelation time given by Eq.~(4.13) (short dashes), and (iii) an 
alternative ${\chi}_{est}$, again based on the `quasi-isotropized' Eq.~(4.9), 
but now allowing for a distribution $N[K]$ generated from time-series data and 
an autocorrelation time (4.15) (long dashes). Both estimates are comparable
in magnitude to ${\chi}_{num}$, but both miss the nontrivial dip that arises
near $E=0.0$.

\section{ESTIMATES OF LYAPUNOV EXPONENTS IN LOWER-DIMENSIONAL HAMILTONIAN
SYSTEMS}
\subsection{ Estimates of the true Lyapunov exponent}
Overall, Eq.~(4.15) first proposed by Casetti {\it et al.}, modified to
allow for moments computed assuming a microcanonical distribution, appears to
give reasonable estimates of the largest Lyapunov exponent in lower-dimensional
Hamiltonian systems, provided that an appreciable fraction of the phase
space corresponds to chaotic orbits. In particular, as long as the true
Lyapunov exponents are not very small (${\chi}_{num}{\not\ll}{\;}t_{D}^{-1}$) 
and/or `stickiness' is not especially prominent, the estimated ${\chi}_{est}$ 
typically agree with the numerical ${\chi}_{num}$ to within factors of two.
In some cases, such as for the {\it FPU} model, the agreement between
${\chi}_{num}$ and ${\chi}_{est}$ rapidly increases with increasing $D$; but
in other cases, such as for the dihedral potential, this is {\it not} the case.
This would suggest that the quasi-isotropy assumption, which should become
increasingly justified in higher dimensions, is {\it not} necessarily the
principal source of error.

FIGURE~\ref{fig-11} compares the numerical and estimated ${\chi}(E)$ for the 
dihedral potential for $D=2$, $3$, $4$, and $6$. The estimated values were
computed using Eq.~(4.13), based on a Gaussian distribution,
with moments generated both from a time-series analysis (dashed lines) and
assuming a microcanonical distribution (dotted lines). The numerical values
are connected with a solid line. One observes significant differences in the
shapes of the curves associated with the numerical and estimated values, but
there is invariably an overall agreement to within a factor of two. The most
striking discrepancies arise for $D=2$, where the estimates completely miss
the abrupt dip in ${\chi}_{num}$ for $E{\;}{\sim}{\;}0$. The fact that, for
$D=2$, the two estimated curves differ significantly at high energies reflects
the fact that the constant energy hypersurface contains large regular islands,
so that the invariant distribution is far from microcanonical.

FIGURE~\ref{fig-12} exhibits the numerical and estimated ${\chi}(E)$ for the
{\it FPU} model for $D=4$, $5$, and $6$, with the estimated values again
computed assuming Gaussian distributions and moments generated from a
time-series analysis. The data have been plotted
on a log-log plot to facilitate comparison with FIG. 3 in Casetti {\it et al}.
Here two points are immediately obvious: (1) The estimated and numerical
curves are distinctly different, with ${\chi}_{est}$ always larger than
${\chi}_{num}$, but their curvatures are comparatively similar. (2) The 
agreement between ${\chi}_{num}$ and ${\chi}_{est}$ becomes progressively 
better for higher dimensions and for higher energies. For $E=5$ in $D=4$, 
where the numerical ${\chi}_{num}=0.211$ corresponds to a very long time
$t{\;}{\sim}{\;}45{\;}{\gg}{\;}t_{D}$, ${\chi}_{est}$ overestimates 
${\chi}_{num}$ by nearly a factor of seven. For $D=6$ and $E=5$, 
${\chi}_{est}$ yields a value approximately $2.65$ times too large; for 
$E=10240$, its value is only $1.27$ times too large.

FIGURE~\ref{fig-13} (a) exhibits the same data for the sixth order truncation
of the Toda potential. As for the {\it FPU} model, ${\chi}_{est}$ 
systematically overestimates the true ${\chi}_{num}$, the largest errors
arising at low energies, where ${\chi}_{num}$ is comparatively small, larger
regular islands exist, and `stickiness' is especially important.

\subsection{ Short-time Lyapunov exponents}
The computations described in the preceding subsection indicate that, for
a variety of lower-dimensional systems, the Casetti {\it et al.} model of a 
stochastic-oscillator equation yields reasonable estimates of the 
largest Lyapunov exponent, ${\chi}$, as a function of energy $E$. However, if
the stochastic-oscillator picture is to be accepted as completely valid, one
must also demand that it `explain' the varying degrees of chaos manifested by
different chaotic orbit segments with the same energy, as probed by short
time Lyapunov exponents. In particular, one might hope that, even if the
estimated values ${\chi}_{est}$ of the true Lyapunov exponent disagree 
significantly with the values ${\chi}_{num}$ computed numerically, the 
estimated and computed values of short-time Lyapunov exponents for different
orbit segments with the same energy should be strongly correlated. For 
example, chaotic segments for which the true short-time ${\chi}_{num}$ is 
especially small should yield estimates ${\chi}_{est}$ that are also
especially small. 

That such correlations do in fact exist is illustrated in FIGS.~\ref{fig-13}
(b) and (c) and FIG.~\ref{fig-14}, which exhibit results appropriate for, 
respectively, the truncated Toda and dihedral potentials. 
Each of these FIGURES was generated by (i) selecting a representative ensemble 
of 1000 initial conditions, all with the same energy; (ii) evolving these into 
the future for a time $T=1024$ while simultaneously tracking the evolution of a 
small perturbation so as to generate ${\chi}_{num}$; (iii) recording the 
values of $K$ for each orbit at fixed intervals ${\delta}t=0.4$; (iv) 
analyzing each orbit to extract ${\langle}K{\rangle}$ and ${\delta}K$; and
(v) using these two moments along with Eq.(4.13) to generate an estimated 
${\chi}_{est}$. The scatter plots provide unambiguous visual confirmation that 
the values of ${\chi}_{num}$ and ${\chi}_{est}$ are strongly related.

This visual impression can be quantified by computing the Spearman rank 
correlation ${\cal R}$ between the values of ${\chi}_{est}$ and ${\chi}_{num}$ 
in each ensemble, which satisfies
\begin{equation}
{\cal R}({\chi}_{num},{\chi}_{est})=1 - {6\over N^{3}-N}\;
\sum_{i=1}^{N}{\delta}_{i}^{2}.
\end{equation}
Here $N=1000$ denotes the number of orbits in the ensemble and ${\delta}_{i}$
denotes the difference in rank for the {\it i\,}th orbit when ordered in
terms of ${\chi}_{num}$ and ${\chi}_{est}$. ${\cal R}=1$ corresponds to a
perfect correlation; 
${\cal R}=-1.0$ corresponds to a complete anti-correlation.

The data sets in FIG.~\ref{fig-13} (b) and (c), corresponding to $E=30$ and
$E=50$ in the truncated Toda potential, both yield ${\cal R}{\;}{\approx}{\;}
0.88$. The data sets in FIG.~\ref{fig-14} (a) - (e), corresponding to $E=1.0$
in the dihedral potential, yield rank correlations ranging between a low of
${\cal R}{\;}{\approx}{\;}0.85$ for $D=3$ and a high of 
${\cal R}{\;}{\approx}{\;}0.95$
for $D=2$. The especially high rank correlation for $D=2$ might seem
surprising since the ensemble contains a large number of regular orbit
segments, with very small ${\chi}_{num}$. The reason ${\cal R}$ remains as
large as it does is that, for orbit segments that are manifestly regular,
so the ${\chi}_{num}$ eventually decays to zero, there is a correlation 
between the estimated value ${\chi}_{est}$ and the rate at which ${\chi}_{num}$
tends to zero: for regular orbits where the short-time ${\chi}_{est}$ is
especially large, the convergence is especially slow, so that, at finite times,
${\chi}_{num}$ will also be especially large\cite{Kan1}.

Not surprisingly, the computed value of ${\cal R}$ for a given ensemble of
initial conditions depends on the total integration time. If the orbits be
integrated for a sufficiently large $T$, their differences tend to `wash
out,' so that the observed range of values for ${\chi}_{num}$ and 
${\chi}_{est}$ both decrease. Eventually, the differences between different
orbit segments become small and the pronounced correlation disappears. 

It would seem visually from FIG.~\ref{fig-14} (a) - (e) that the numerical
and estimated values of the short-time Lyapunov exponents deviate largely
because of some overall scaling. Given that at least for $D{\;}{\ge}{\;}3$, 
the phase space is almost entirely chaotic, so that the distribution of 
curvatures reflects a microcanonical distribution, the evidence (cf. 
FIG.~\ref{fig-6}) that this implies a nearly Gaussian distribution, and the 
argument in the preceding subsection that quasi-isotropy is not necessarily 
the principal source of discrepancies, it would seem natural to conjecture that
this reflects a misidentification of the proper autocorrelation
time ${\tau}$. Panel (f) in FIG.~\ref{fig-14} shows what happens to the 
estimated value ${\chi}_{est}$ for the $D=6$ dihedral potential if, for each
orbit, ${\tau}$ is reduced by a factor of ${\approx}{\;}0.75$, the value 
required to ensure that, for the ensemble, the mean values of the estimated 
and numerical exponents coincide, {\it i.e.,} 
${\langle}{\chi}_{num}{\rangle}={\langle}{\chi}_{num}{\rangle}$. The net
result is that, to a fair degree of approximation, the data points are aligned
along ${\chi}_{est}={\chi}_{num}$.

\subsection{ The special case $D=2$}
It is natural to ask whether one can relax the assumption of quasi-isotropy,
at least for $D=2$ where it would seem most suspect. One way in which to do 
this would be to work instead with the Jacobi metric, which, for $D=2$, leads 
to a single oscillator equation of the form\cite{Kan}\cite{CSP}
\begin{equation}
{d^{2}{\xi}\over dt^{2}}-{1\over W}{dW\over dt}{d{\xi}\over dt}+{\tilde K}
{\xi}=0,
\end{equation}
where $W=E-V(q^{i})$ denotes the kinetic energy and
\begin{equation}
{\tilde K}={\Delta}V+{1\over 2W}|{\nabla}V|^{2}.
\end{equation}
Unfortunately, however, this equation is very difficult to explore numerically
since it yields near-divergences for $W{\;}{\approx}{\;}0$, which prove quite 
common for $D=2$. 

Alternatively, within the setting discussed hitherto in this paper, there
are two ways in which one might proceed:
\par\noindent
1. Consider the full multi-dimensional Jacobi equation and view it as a
matrix stochastic equation. This could at least provide the `average' rate of
exponential instability in different configuration space directions. Quite 
generally, a small perturbation ${\xi}^{i}$ will satisfy
\begin{equation}
{d^{2}{\xi}^{i}\over dt^{2}}+\sum_{j}V_{ij}{\xi}^{j}=0,
\end{equation}
with $V_{ij}{\;}{\equiv}{\;}{\partial}^{2}V/{\partial}q^{i}{\partial}q^{j}$
the second derivative matrix. The objective then is to view each component 
of this matrix as an (approximately) independent stochastic variable,
{\it i.e.,} considering
\begin{equation}
V_{ij}=V_{ji}={\Omega}_{0,ij}+{\delta}{\Omega}_{ij},
\end{equation} 
with ${\delta}{\Omega}_{ij}$ a random variable. 
Given distributions $N[V_{xx}]$, $N[V_{yy}]$, and $N[V_{xy}]=N[V_{yx}]$, which 
can be computed from time-series data or assuming ergodicity, and some 
approximation to the autocorrelation functions ${\Gamma}_{xx}$, 
${\Gamma}_{yy}$, and ${\Gamma}_{xy}$, which can again be motivated either 
from a time series or assuming ergodicity, this system is easy to solve 
numerically.\cite{Kan3}
\par\noindent
2. At least for $D=2$, it is easy to diagonalise the matrix equation (4.2)
at any given instant so as to obtain the eigenvalues of the stability matrix.
The corresponding eigenvectors will then satisfy equations of the form
\begin{equation}
{d^{2}{\xi}_{\pm}\over dt^{2}}+{\omega}_{\pm}{\xi}_{\pm} = 0,
\end{equation}
where the time-independent eigenvalues satisfy
\begin{equation}
{\omega}_{\pm}={1\over 2}\left( V_{xx}+V_{yy}\right) {\pm}{1\over 2}
\left[ \left( V_{xx}-V_{yy}\right)^{2} + 4V_{xy}^{2} \right]^{1/2}.
\end{equation}
Viewing ${\omega}_{\pm}={\Omega}_{0,\pm}+{\delta}{\Omega}_{\pm}$ as stochastic 
variables leads to a pair of
decoupled oscillator equations which are easy to solve numerically.
In general, one might anticipate that the smaller eigenvalue will correspond
to the more rapid growth rate.

\begin{figure}[t]
\centering
\centerline{
        \epsfxsize=8cm
        \epsffile{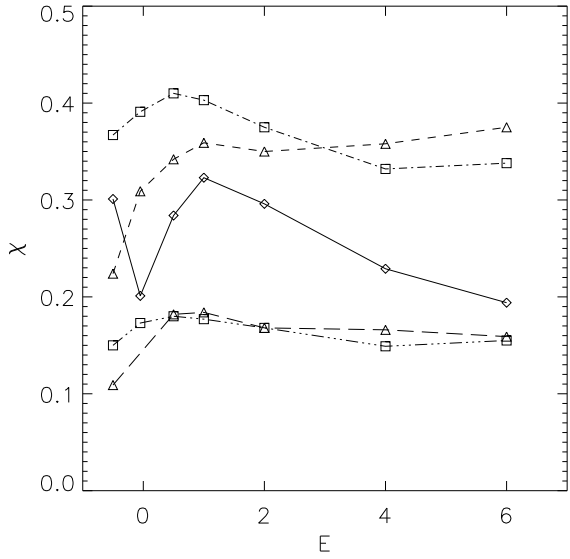}
           }
        \begin{minipage}{10cm}
        \end{minipage}
        \vskip -0.0in\hskip -0.0in
\caption{\label{fig-10} Estimated values of the largest Lyapunov exponent
for the $D=2$ dihedral potential as a function of energy: ${\chi}_{num}$
generated from direct numerical integration (solid curve), the Casetti 
{\it et al.} value, generated assuming a Gaussian $N[K]$ and autocorrelation
time ${\tau}$ given by Eq.~(4.12) (short dashes); an estimate based on 
Eq.~(4.8), but now using the $N[K]$ generated from a time-series analysis and 
${\tau}$ given by Eq.~(4.14) (long dashes); an estimate based on
Eq.~(6.6), using ${\omega}_{-}$ and ${\tau}$ given by Eq.~(4.14) (dot-dashes);
and an estimate based on the coupled oscillator system, assuming Gaussian
fluctuations and ${\tau}$ given by Eq.~(4.12).}
\vspace{-0.0cm}
\end{figure}

\begin{figure}[t]
\centering
\centerline{
        \epsfxsize=8cm
        \epsffile{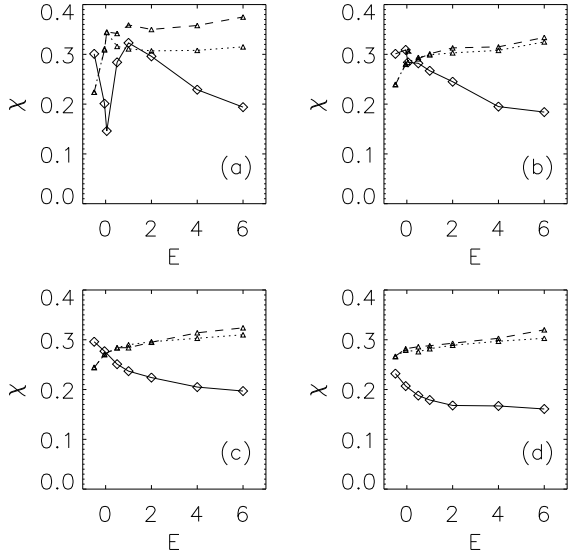}
           }
        \begin{minipage}{10cm}
        \end{minipage}
        \vskip -0.0in\hskip -0.0in
\caption{\label{fig-11}
Estimated values of Lyapunov exponents for orbits in the dihedral potential,
generated from numerical integrations (solid lines) and estimated using 
Eq.~(4.12). (a) For $D=2$. (b) $D=3$. (c) $D=4$. (d) $D=6$.}
\vspace{-0.0cm}
\end{figure}

\begin{figure}[t]
\centering
\centerline{
        \epsfxsize=8cm
        \epsffile{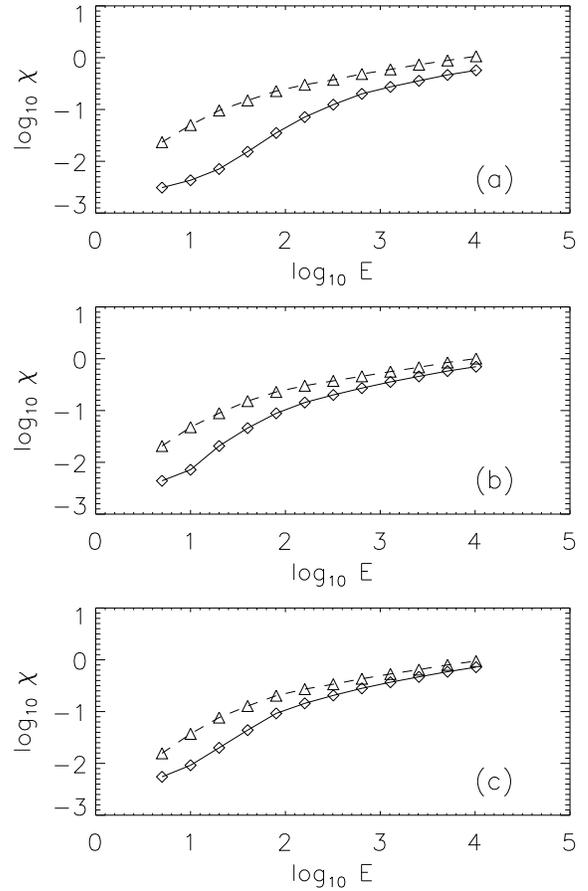}
           }
        \begin{minipage}{10cm}
        \end{minipage}
        \vskip -0.0in\hskip -0.0in
\caption{\label{fig-12}
Estimated values of Lyapunov exponents for orbits in the {\it FPU} model,
generated from numerical integrations (solid lines) and estimated using 
Eq.~(4.12). (a) For $D=4$. (b) $D=5$. (c) $D=6$.}
\vspace{-0.0cm}
\end{figure}

\begin{figure}[t]
\centering
\centerline{
        \epsfxsize=8cm
        \epsffile{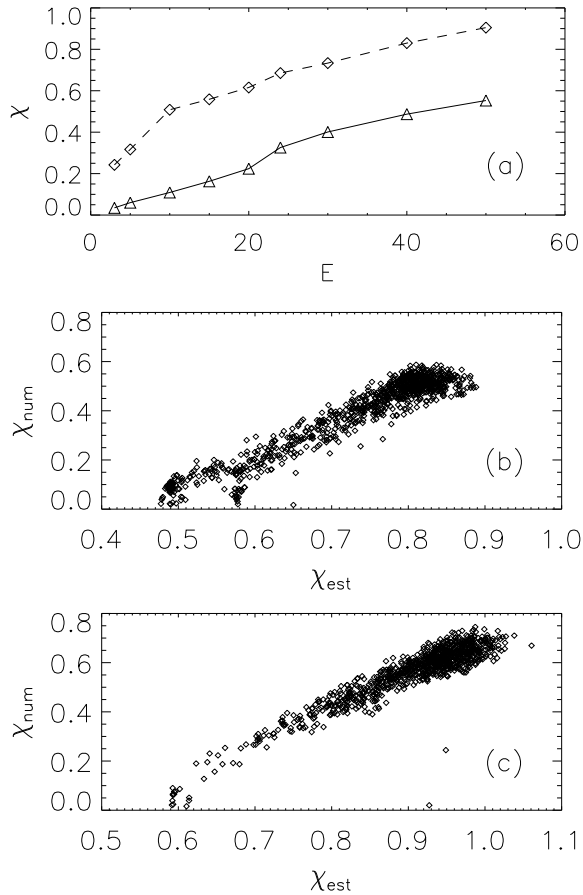}
           }
        \begin{minipage}{10cm}
        \end{minipage}
        \vskip -0.0in\hskip -0.0in
\caption{\label{fig-13}
(a) Estimated values of Lyapunov exponents for orbits in the truncated
Toda potential,
generated from numerical integrations (solid lines) and estimated using 
Eq.~(4.12) (b) Short-time lyapunov exponents computed using Eq.~(4.12) 
(${\chi}_{est}$) and generated from numerical integrations (${\chi}_{num}$)
for $E=30.0$. (c) The same for $E=50.0$.}
\vspace{-0.0cm}
\end{figure}

\begin{figure}[t]
\centering
\centerline{
        \epsfxsize=8cm
        \epsffile{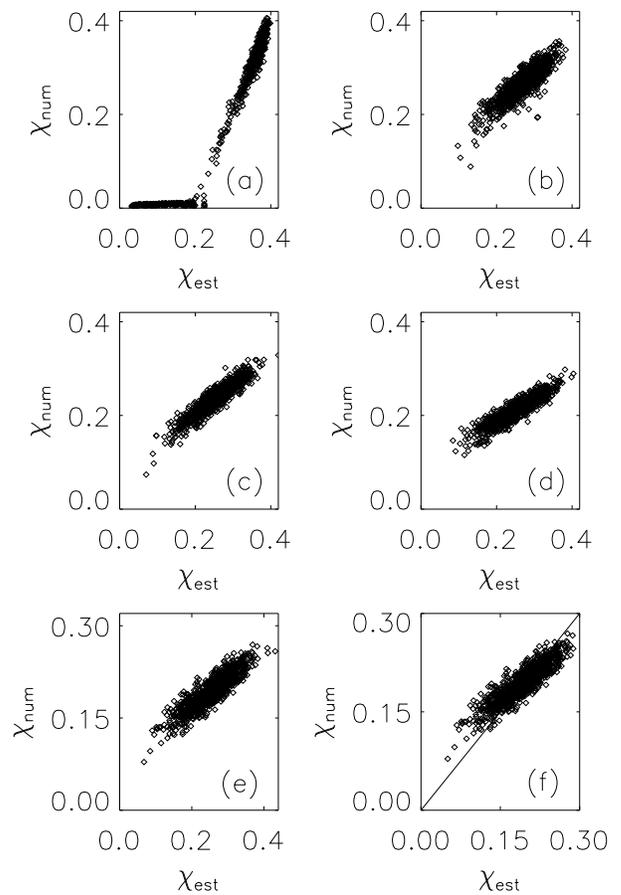}
           }
        \begin{minipage}{10cm}
        \end{minipage}
        \vskip -0.0in\hskip -0.0in
\caption{\label{fig-14}
(a) Short-time lyapunov exponents computed using Eq.~(4.12) (${\chi}_{est}$) 
and generated from numerical integrations (${\chi}_{num}$) for $E=1.0$ in the
$D=2$ dihedral potential. (b) The same for $D=3$. (c) $D=4$. (d) $D=5$.
(e) $D=6$. (f) ${\chi}_{num}$ for $D=6$ contrasted with revised estimates 
${\chi}_{est}$
generated by rescaling the time scale ${\tau}$ of Eq.~(4.12) by a factor of
$0.75$.}
\vspace{-0.0cm}
\end{figure}

These alternative were tested in detail for the $D=2$ dihedral potential.
The principal results are summarised in FIG.~\ref{fig-10}, which shows the 
numerical ${\chi}_{num}(E)$ (solid line) as well as estimated values 
${\chi}_{est}(E)$ 
generated in four different ways. The short and long dashed lines, discussed
already in the preceding section, correspond to the isotropized equation (4.9),
assuming a microcanonical distribution (short dashes) or using inputs generated
from orbital data (long dashes). The triple-dot dashed curve represents the 
values generated for the coupled oscillator system and the dot dashed curve
represents the values generated by solving Eq.~(6.6) for ${\omega}_{-}$.
All the estimated curves yield values ${\chi}_{est}$ that agree with 
${\chi}_{num}$ to within factor of two, but none seems especially better 
than the others. 

The hypothesis that chaotic behavior in lower-dimensional Hamiltonian systems 
can be modeled by a stochastic-oscillator equation would appear robust in the 
sense that different implementations all lead to predictions that
yield at least rough agreement with numerical integrations. However, none
of the alternatives considered here would appear `completely right.' It
seems likely that, in very low-dimensional systems, the details matter
sufficiently that no universal prescription will yield a completely 
accurate prediction.

\section{CONCLUSIONS AND DISCUSSION}
The results described in this paper strongly corroborate the intuition that
chaotic motion in lower-dimensional Hamiltonian systems can be visualized as 
random, so that the average instability of chaotic orbits, and hence the 
values of their largest Lyapunov exponents, can be derived from a harmonic 
oscillator equation with a randomly varying frequency.
Modulo straightforward modifications, technical rather than conceptual in
nature, the approach introduced by Casetti {\it et al.} for higher-dimensional
systems also works reasonably well for systems with dimensionality as low as 
$D=2$. In this sense, as suggested in the Introduction, one would appear to
have a clear new paradigm in terms of which to interpret the origins of chaos 
in lower-dimensional Hamiltonian systems. 

The precise numerical values of ${\chi}$ predicted using this analytic 
approach are somewhat less accurate in lower dimensions than they are
for much larger $D$, but it remains true that, in general, this approach
yields predictions that are correct to within factors of two. In principle
one might hope to do still better but, as a practical matter, this would 
seem difficult if not impossible. The `obvious' alternatives considered in
Section VI C yield somewhat different predictions for the largest Lyapunov
exponents. In some cases these predictions are somewhat better than those
based on Eqs.~(4.12), (4.15), and (4.16) but, overall, they seem neither 
appreciably better nor appreciably worse. This would suggest that the 
predictions based on these equations are comparatively robust, in the sense 
that minor modifications do not yield vastly different results. However, 
this might also suggest that there is no single, universal modification that 
one might introduce which would yield near-perfect agreement for all 
potentials and energies. In point of fact, this is hardly surprising. There 
is every reason to expect that details which should `wash out' in 
higher-dimensional systems will remain important in lower-dimensional
systems. A `thermodynamic' description of chaos should work best for systems
with many degrees of freedom.

In this context, two significant points should be stressed. 

(1) Even when the predicted values $\chi_{est}(E)$ of the `true' Lyapunov 
exponent ${\chi}_{num}(E)$ are off by as much as factors of two, one observes 
strong correlations between $\chi_{est}(E)$ and ${\chi}_{num}(E)$ for 
different orbit segments with the same energy. Orbit segments for which the 
predicted 
${\chi}_{est}$ is low tend to have small short-time exponents ${\chi}_{num}$;
and, similarly, segments for which ${\chi}_{est}$ is high tend to have
larger values of ${\chi}_{num}$. The physics entering into Eqs.~(4.12),
(4.15), and (4.16) allows one to distinguish clearly between orbit segments
that are `wildly chaotic' in visual appearance and have especially large
short-time exponents and `sticky' segments that are nearly regular in
appearance and have comparatively small short-time exponents. 

(2) The largest discrepancies between the predicted and numerically
computed Lyapunov exponents occur invariably for those potentials and
energies where large portions of the chaotic sea correspond to `sticky'
orbits manifesting nearly regular behavior, in which case ${\chi}_{est}$
can be much larger than the `true' ${\chi}_{num}$. This is exactly what one
would expect. If large portions of the stochastic sea are `sticky,' an
orbit will spend much of its time evolving in a nearly regular fashion,
but it is clear that, while manifesting such near-regular behavior, its
motion cannot be characterised as essentially random. Indeed, as discussed 
more carefully elsewhere~\cite{Kan}, chaotic orbit segments for which 
${\langle}K{\rangle}$ and ${\delta}K$ assume values close to those 
characteristic of regular orbits tend systematically to have very small
short-time Lyapunov exponents.

The principal difference between the approach developed in this paper and
the approach introduced by Casetti {\it et al.} is that the statistical
properties of the mean curvature $K$ are not derived assuming a canonical
distribution. For a truly conservative system, a thermodynamic description
must, strictly speaking, be based on a microcanonical distribution, and it
is only for large $D$ that one can approximate such a `correct' description
by a description based on a canonical distribution. Moreover, for very low
dimensions, especially $D=2$, even a microcanonical distribution is clearly
unjustified. A microcanonical analysis is based on the assumption that the
entire constant-energy hypersurface is chaotic, but for lower dimensions,
non-integrable systems typically exhibit a coexistence of regular and chaotic
regions, both with significant measure. A correct analysis must involve
deriving the statistics of the curvature only in the chaotic phase-space 
regions, a task which seems difficult analytically but, given an assumption
of ergodicity, is straightforward to implement via a time-series analysis.

In part, this work concerning chaos and the phase mixing of chaotic orbits 
was motivated in the context of nonequilibrium systems comprised of a large
number of interacting particles.  Examples of such systems include
self-gravitating systems, {\it e.g.}, galaxies, and charged-particle
beams governed by external focusing forces and internal Coulomb forces
(space charge).  For both these examples fast evolutionary time scales have
profound consequences.  For galaxies, they are an integral part of the
formation process~\cite{kan98}.  For beams, they limit the degree to which an
accelerator designer can preserve the beam quality, especially insofar
as the evolution is irreversible~\cite{bohn00}.

As mentioned in Section II, one way to infer the fastest time scale is to 
consider the interaction of a single particle with the coarse-grained 
potential formed by all the other particles.  The problem then reduces to 
one involving a low-dimensional Hamiltonian, and the obvious question is
to what extent statistical arguments concerning the behavior of chaotic 
single-particle orbits can be invoked to simplify the analysis further.  All 
the examples presented herein suggest that time scales in low-dimensional 
Hamiltonian systems inferred via the statistical methods of Casetti 
{\it et al.}~are typically valid within a factor of order two.  They also 
suggest that uncertainties in the computation of these time scales are 
principally associated with uncertainties in the autocorrelation time, and 
that these time scales are comparatively insensitive to the choice of the 
invariant measure that weights the statistical averages.  More importantly, 
however, our examples reinforce the idea advanced by Cerruti-Sola 
and Pettini~\cite{csp96} that rapid mixing originates from parametric 
instability due to positive-curvature fluctuations along the geodesic 
trajectories of the particles over the configuration-space manifold.  
Cerruti-Sola and Pettini conjectured that: ``This mechanism is apparently 
the most relevant -- and in many cases unique -- source of chaoticity in 
physically meaningful Hamiltonians."  The diverse set of examples presented
here would seem to corroborate their conjecture.

The statistical analysis, however, does carry some {\it caveats}.  It 
generally `predicts' fast exponential mixing in potentials that are 
known {\it a priori} to be integrable and thereby admit only regular 
orbits that can only mix secularly.  Examples include spherically 
symmetric potentials for which Poisson's equation generates nonuniform 
density distributions, and special triaxial potentials such as the Staeckel
potentials~\cite{deZeeuw85}.  Thus the analysis provides no information 
as to what criteria are necessary and sufficient to establish a 
preponderance of globally chaotic orbits; it merely hypothesizes their 
existence.  Related to this deficiency is the failure of the analysis to 
account for `sticky' chaotic orbit segments that, when present, will 
tend to slow down the mixing.  Real systems may, however, mitigate these 
{\it caveats}.  For example, external noise, even with very small 
amplitude, is known to add greatly to the efficiency of chaotic mixing 
by overcoming `stickiness'~\cite{siop00}.  And the presence of 
localized irregularities that have been coarse-grained away may increase 
the chaoticity of the orbits.  An important point, however, is that 
graininess which manifests itself in binary particle interactions is 
{\it not} an example of such localized irregularities.  Graininess 
establishes diffusion of an orbit away from the trajectory it would have 
in the smooth potential but, at least for nonchaotic orbits, this diffusion 
involves a secular, rather than
exponential, divergence of trajectories~\cite{sideris01}.

Because it is based on the Eisenhart metric, the present treatment is 
restricted to stationary systems.  However, with a Finsler metric, the 
geometric method can also incorporate potentials that are explicitly 
time-dependent and/or velocity-dependent~\cite{dibari97}.  For example, 
recent work involving the H\'{e}non-Heiles potential~\cite{cipriani982} 
resulted in a geometric measure of chaos over the associated Finsler 
manifold that was used for fast computation of the system's Poincar\'{e} 
surface of section.  If used with a coarse-grained potential, the 
Eisenhart metric includes no mechanism for changing the particle 
energies.  In principle, it can be included with a Finsler metric based 
on a time-dependent coarse-grained potential; however, the 
generalization also requires specifying a suitable invariant measure for 
the nonequilibrium system~\cite{gallavotti00}.

One final point should be noted. In writing this paper, the authors have
deliberately adopted a tact somewhat complementary to that adopted by 
Casetti {\it et al}. Rather than focusing on the differential geometry of
spaces admitting an Eisenhart metric, the discussion has, to the extent
possible, been couched in the language of conventional Hamiltonian mechanics.
Such an approach serves to make the physical ideas underlying the general
approach more transparent physically and, it is hoped, will make the picture
of chaotic motion as a random process comprehensible to a substantially
larger audience.

\acknowledgments
This research was supported in part by NSF AST-0070809, and in part by the Universities Research Association, Inc., under contract DE-AC02-76CH00300 with the U.S. Department of Energy.

\pagestyle{empty}

\pagestyle{empty}

\pagestyle{empty}
\vfill\eject
\pagestyle{empty}

\pagestyle{empty}

\pagestyle{empty}

\pagestyle{empty}
\vfill\eject
\pagestyle{empty}

\pagestyle{empty}
\vfil\eject
\pagestyle{empty}

\pagestyle{empty}
\vfill\eject
\pagestyle{empty}

\pagestyle{empty}

\pagestyle{empty}


\begin{references}
\bibitem{Kan2} Recall that, at least for a Gaussian distribution, the form of 
${\Gamma}$ is known to be irrelevant: any ${\Gamma}$ that yields the correct
autocorrelation time will yield the same estimated ${\chi}_{est}$.
\bibitem{Kan3}Unfortunately, an analytic solution using the approach discussed
in Section IV seems impossible. For a single oscillator equation, one is
led to three coupled moment equations and, hence, a cubic equation for ${\chi}$.
For two coupled oscillators, one has eight coupled equations and, hence, an
eighth order equation for ${\chi}$.
\end{references}
\end{document}